\documentclass[3p,times,preprint]{elsarticle}

\usepackage{amssymb}
\usepackage{lipsum}

\usepackage[colorlinks=true,linkcolor=blue,citecolor=blue,urlcolor=blue]{hyperref}

\newif\ifdraft

\ifdraft
  \graphicspath{{figures_low/}}
\else
  \graphicspath{{figures/}}
\fi

\usepackage{lineno}

\journal{Nuclear Physics A}

\begin{document}

\begin{frontmatter}



\title{Effects of irradiation by protons, neutrons, and gamma particles on electrical properties of 4H-SiC diodes and LGAD sensors}


\author[first]{Jiří Kroll}
\author[first]{Pavla Federičová}
\author[second]{Jan Chochol}
\author[second]{Adam Klimsza}
\author[first]{Jana Kozáková}
\author[second]{Adam Kozelský}
\author[third]{Vojtěch Kráčmar}
\author[first]{Jiří Kvasnička}
\author[second]{Roman Malousek}
\author[third]{Mária Marčišovská}
\author[third]{Michal Marčišovský}
\author[first]{Marcela Mikeštíková}
\author[second]{David Novák}
\author[third]{Radek Novotný}
\author[second]{Peter Slovák}
\author[second]{Radim Špetík}
\author[first,third]{Peter Švihra}
\author[first]{Pavel Tůma}

\affiliation[first]{organization={Institute of Physics of the Czech Academy of Sciences},
            addressline={Na Slovance 2}, 
            city={Prague 8},
            postcode={18200}, 
            country={Czechia}}

\affiliation[second]{organization={onsemi},
            addressline={1. maje 2230}, 
            city={Rožnov pod Radhoštěm},
            postcode={75661}, 
            country={Czechia}}

\affiliation[third]{organization={Faculty of Nuclear Sciences and Physical Engineering, Czech Technical University in Prague},
            addressline={Brehova 78/7}, 
            city={Prague},
            postcode={11519}, 
            country={Czechia}}

\begin{abstract}
4H-SiC is a wide-bandgap semiconductor with high displacement threshold energy, large critical electric field, and low intrinsic carrier concentration, making it attractive for radiation-hard detector applications. In this work, we investigate the electrical characteristics of 4H-SiC \mbox{P$^{+}$-in-N} (PN) diodes and Low-Gain Avalanche Detectors (LGADs) fabricated by onsemi before and after irradiation by \mbox{24 GeV/c} protons, reactor neutrons, and $^{60}$Co gamma rays. Current–voltage (IV) and capacitance–voltage (CV) measurements were performed at room temperature for proton fluences up to $1\times10^{16}\;\mathrm{protons/cm^2}$, neutron fluences up to $1\times10^{18}\;\mathrm{1\;MeV\;n_{eq}/cm^2}$, and total ionizing doses up to \mbox{300 kGy}.

Hadron irradiation induces pronounced changes in both leakage current and bulk capacitance, consistent with radiation-induced formation of deep acceptor-like defects and strong compensation of the originally N-type material. For high proton fluences, the leakage current decreases and the bulk capacitance becomes bias-independent, indicating effective compensation of the epitaxial layer. Extreme neutron fluences lead to a substantial expansion of the depleted region into the originally highly doped substrate, as inferred from the measured capacitance values. Gamma irradiation up to \mbox{300 kGy} results in significantly modified capacitance behavior, suggesting reduction of the effective doping concentration in the epitaxial and multiplication layers.

The results demonstrate that radiation-induced compensation strongly modifies the effective space charge in 4H-SiC devices at high hadron fluences, while the leakage current is influenced by the enlarged depletion volume together with field-enhanced and surface-related generation mechanisms. In contrast, ionizing damage primarily affects the effective doping and electric-field distribution.  
\end{abstract}



\end{frontmatter}




\section{Introduction}
\label{introduction}
\noindent Silicon carbide (SiC) is a wide-bandgap semiconductor material available in several polytypes, among which 4H-SiC is the most technologically mature for electronic device fabrication. Owing to its large bandgap (3.26 eV), high critical electric field, high carrier saturation velocity ($1.47\times10^7\;\mathrm{cm/s}$ for electrons and $0.69\times10^7\;\mathrm{cm/s}$ for holes)~\cite{gsponer2025}, and high atomic displacement threshold energies (21 eV for C and 35 eV for Si) \cite{rafi2020,moscatelli2006}, 4H-SiC exhibits intrinsic properties favorable for operation in high-radiation and high-temperature environments. In addition, its intrinsic carrier concentration at room temperature is approximately 18–19 orders of magnitude lower than in silicon, which results in very low thermally generated leakage currents even after substantial radiation exposure.

These characteristics make 4H-SiC a promising candidate for radiation-hard particle detectors. However, the large bandgap also implies a relatively low number of generated electron–hole pairs per unit path length (approximately 57 e--h pairs per $\mu$m for a minimum ionizing particle). At the same time, currently achievable epitaxial thicknesses typically do not exceed about 100~$\mu$m. For applications requiring large signal amplitude, such as precision timing or tracking, internal charge multiplication is therefore desirable. This concept is well established in silicon Low-Gain Avalanche Detectors (LGADs), where a highly doped gain layer enables controlled avalanche multiplication \cite{pellegrini2014}.

Recent advances in industrial 4H-SiC wafer production, driven primarily by the power electronics market, have significantly improved material quality and device reproducibility. First-generation 4H-SiC LGAD structures have already demonstrated internal gain and promising performance \cite{novotny2025,svihra2025}. Nevertheless, systematic studies addressing their behavior under high hadron fluences and extreme neutron irradiation remain limited, particularly with respect to compensation effects and evolution of the effective depletion region.  

\section{Design of 4H-SiC PN and LGAD sensors}
\label{design}
\noindent The investigated 4H-SiC PN and LGAD diodes are designed as square devices with an active area of $3 \times 3$~mm$^{2}$. The top side is either fully covered by a metal contact with a circular opening of 100~$\mu$m diameter in the center, or by a metal grill consisting of 5~$\mu$m wide strips with a pitch of 15~$\mu$m, enabling laser-based characterization in both configurations \cite{novotny2025}. The devices were fabricated on 6-inch wafers, each containing PN and LGAD structures with both full and grill top metallization.

\begin{figure}
	\centering 
	\includegraphics[width=0.7\textwidth, angle=0]{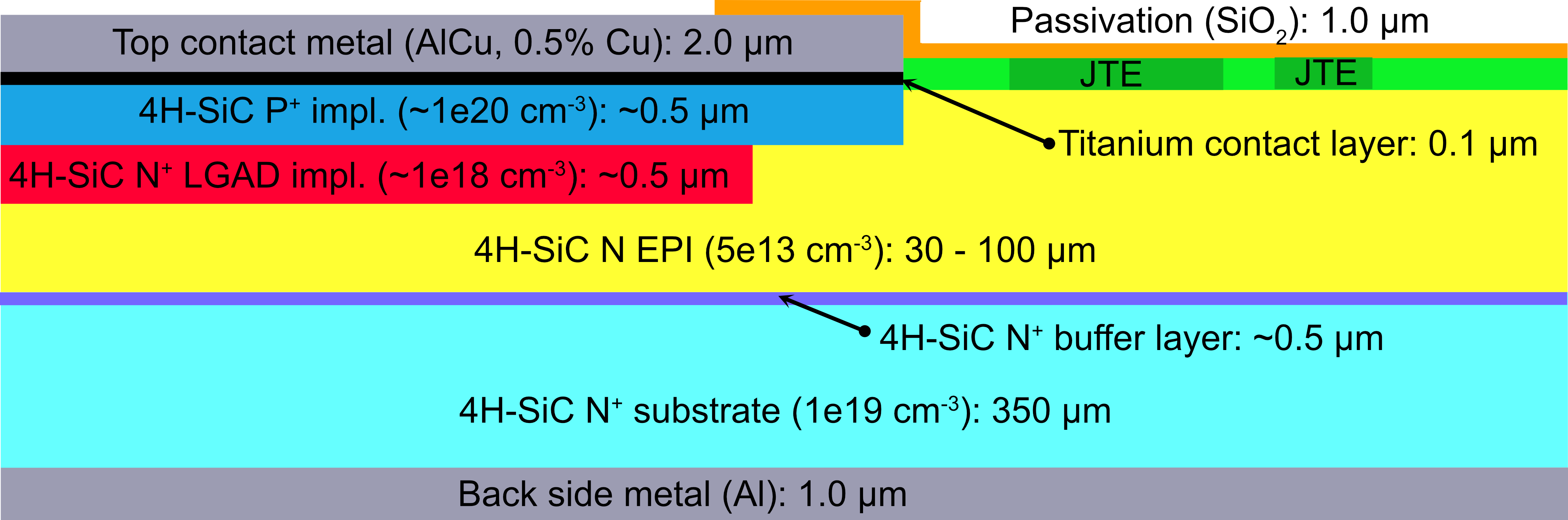}	
	\caption{Schematic cross-sectional view of the 4H-SiC LGAD structure illustrating the highly doped substrate, N-type epitaxial layer, internal multiplication layer, P$^{+}$ implant, metal contacts, and the junction termination extension (JTE).}
	\label{fig:sic_scheme}
\end{figure}

Each wafer consists of a highly doped N-type 4H-SiC substrate with an effective doping concentration of approximately $10^{19}$~cm$^{-3}$, as illustrated in Fig.~\ref{fig:sic_scheme}. On top of the substrate, a high-resistivity N-type epitaxial layer with a thickness between 30 and 100~$\mu$m and a nominal doping concentration of approximately $5 \times 10^{13}$~cm$^{-3}$ is grown. The epitaxial layer is separated from the substrate by a highly doped N-type buffer layer with a thickness of approximately $0.5~\mu$m.

The LGAD multiplication layer, P$^{+}$ implant, and junction termination extension (JTE) are formed by ion implantation. The multiplication layer is created by nitrogen implantation with energies in the range 950--1400~keV and doses between $1.65$ and $1.75 \times 10^{13}$~cm$^{-2}$, depending on the targeted gain; the resulting doping concentration can reach $10^{18}$~cm$^{-3}$. The P$^{+}$ implant and JTE regions are realized by aluminum implantation with energies of 30--200~keV and doses of $1.6$--$8 \times 10^{14}$~cm$^{-2}$ for the P$^{+}$ region, and 300--330~keV with doses of $1.4$--$1.6 \times 10^{13}$~cm$^{-2}$ for the JTE. The peak doping concentration of the P$^{+}$ implant reaches values up to $10^{20}$~cm$^{-3}$.

The backside metallization is formed by aluminum, while the top contact consists of aluminum with a small copper admixture. Ohmic contact to the P$^{+}$ implant is ensured by a 0.1~$\mu$m thick titanium interlayer. The top surface is passivated by a 1.0~$\mu$m thick SiO$_2$ layer.     

\section{Irradiation of 4H-SiC samples}
\label{irradiation}
\noindent After fabrication of the wafers, individual PN, LGAD1, and LGAD2 devices were diced from wafer W1, including both full and grill top metallization variants. The difference between LGAD1 and LGAD2 devices lies in the implantation energy of the nitrogen ions forming the multiplication layer, which determines the achievable internal gain. The diced devices were electrically characterized prior to irradiation, and selected samples were subsequently exposed to proton, neutron, and gamma irradiation.

\subsection*{Proton irradiation}
\noindent Proton irradiation was performed at the CERN IRRAD facility using 24~GeV/c protons delivered by the Proton Synchrotron accelerator \cite{ravotti2015}. The samples were mounted in lightweight holders made of 0.8~mm thick PEEK (polyether ether ketone) with dedicated windows for device placement. Each sample was fixed using non-adhesive Kapton foil covered by Kapton tape, resulting in a total Kapton thickness below 200~$\mu$m per holder. Each PEEK window accommodated up to two 4H-SiC devices. Eleven PEEK holders were aligned sequentially along the beam direction and irradiated simultaneously. Two additional PEEK holders contained thin aluminum foils with dimensions equivalent to the 4H-SiC samples, serving as dosimetry cross-check monitors. One was positioned upstream and one downstream of the sample stack. The total amount of material in the beam was minimized to suppress secondary particle production and avoid artificial enhancement of the delivered fluence \cite{cindro2026}. Irradiation was carried out at ambient temperature (approximately 20$^{\circ}$C) to eight total proton fluences: $5 \times 10^{12}$, $1 \times 10^{13}$, $5 \times 10^{13}$, $1 \times 10^{14}$, $5 \times 10^{14}$, $1 \times 10^{15}$, $5 \times 10^{15}$, and $1 \times 10^{16}$~p/cm$^{2}$.

\subsection*{Neutron irradiation}
\noindent Neutron irradiation was performed at the TRIGA Mark II reactor of the Jožef Stefan Institute in Ljubljana \cite{ambrozic2017}. For each target fluence, several PN, LGAD1, and LGAD2 samples were wrapped in thin aluminum foil, forming three separate packages (one per device type). These three packages were subsequently enclosed together in an additional aluminum foil. The samples were inserted into aluminum cylinders with a diameter of 2.4~cm and length of 10~cm, accommodating devices from multiple user groups, and positioned in the reactor core using dedicated irradiation channels. Three sets of devices were irradiated to total neutron fluences of $2.3 \times 10^{17}$~n$_{\mathrm{eq}}$/cm$^{2}$ (channel F19), $5.2 \times 10^{17}$~n$_{\mathrm{eq}}$/cm$^{2}$ (channel TIC), and $1.0 \times 10^{18}$~1~MeV~n$_{\mathrm{eq}}$/cm$^{2}$ (channel CC). 

\subsection*{Gamma irradiation}
\noindent Gamma irradiation was performed using a $^{60}$Co source at UJP PRAHA a.s. \cite{ujp2026}. The samples were placed in a charged-particle equilibrium box made of aluminum and lead, following ESA/SCC recommendations \cite{cpe2003}. This setup minimizes dose enhancement by establishing electron equilibrium, ensures uniform energy deposition, and suppresses contributions from low-energy scattered radiation. Devices of all three types were irradiated to total ionizing doses (TID) of 0.3, 3, 30, 150, and 300~kGy, with five samples per type and per dose. In the present study, only samples irradiated to 3, 30, and 300~kGy were electrically characterized.   

\subsection*{Hardness factors and post-irradiation handling}
\noindent The non-ionizing energy loss (NIEL) hardness factors in silicon relative to 1~MeV neutrons are 0.62 for 24~GeV/c protons at CERN IRRAD \cite{allport2019} and 0.9 for reactor neutrons at JSI \cite{jsi_hardness_2026}. The neutron fluences reported above are already scaled to 1~MeV neutron equivalent fluence (1~MeV~n$_{\mathrm{eq}}$/cm$^{2}$). In the following text, the unit n$_{\mathrm{eq}}$/cm$^{2}$ always refers to 1~MeV neutron equivalent fluence. For $^{60}$Co gamma irradiation, conversion between TID and 1~MeV neutron equivalent fluence is sample-dependent. For silicon diodes, an approximate relation of 1~MGy $\approx 2 \times 10^{11}$~1~MeV~n$_{\mathrm{eq}}$/cm$^{2}$ has been reported \cite{mikestik2024}. After proton and gamma irradiation, the samples were immediately stored in a freezer at temperatures below $-20^{\circ}$C. Neutron-irradiated devices remained at room temperature for several days to allow for radioactive decay before being transferred to the freezer. Transport from CERN and JSI to Prague was performed at temperatures below $0^{\circ}$C to minimize unintended annealing.

\section{Experimental method}
\label{experiment}
The IV and CV characteristics were measured using a TESLA200 probe station equipped with a shielded triaxial chuck and triaxial probes. The sample under test was placed on the chuck, whose temperature was stabilized at $+20^{\circ}$C, and fixed by vacuum. The top metal contact was accessed using a triaxial probe needle. During the measurements, the chuck was enclosed in an environmental chamber in which the relative humidity was reduced below 1\% by continuous nitrogen flow. The bias voltage was supplied by a Keithley 2657A Source Measure Unit (SMU). The HI HV terminal was connected to the chuck, while the LO HV terminal was connected to the probe. A protective 1~M$\Omega$ resistor was included in the high-voltage line. The leakage current was measured directly by the SMU, with the LO HV terminal additionally referenced to laboratory ground.

\begin{figure*}[t]
  \centering
  \includegraphics[width=0.49\textwidth]{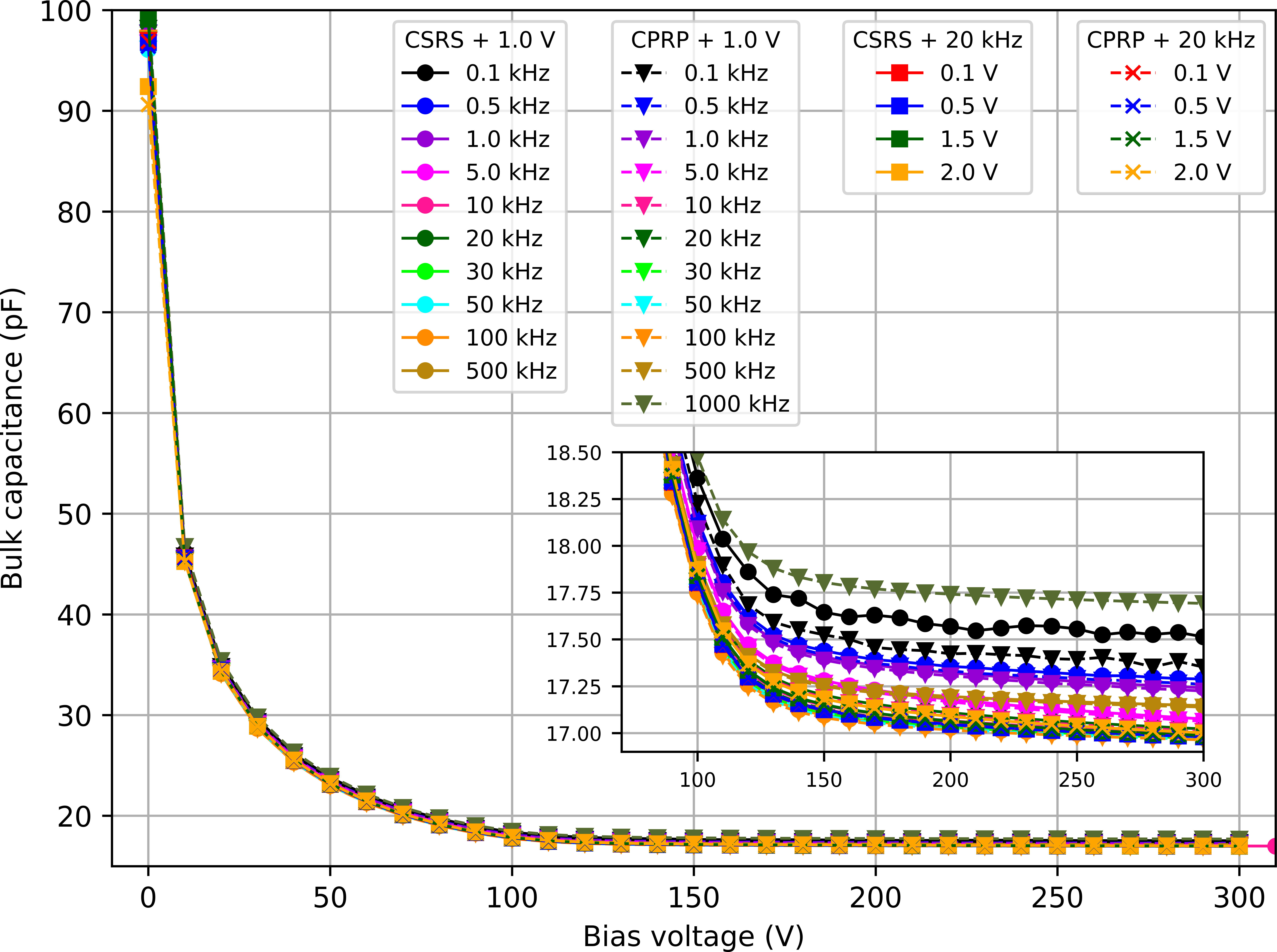}\hfill
  \includegraphics[width=0.49\textwidth]{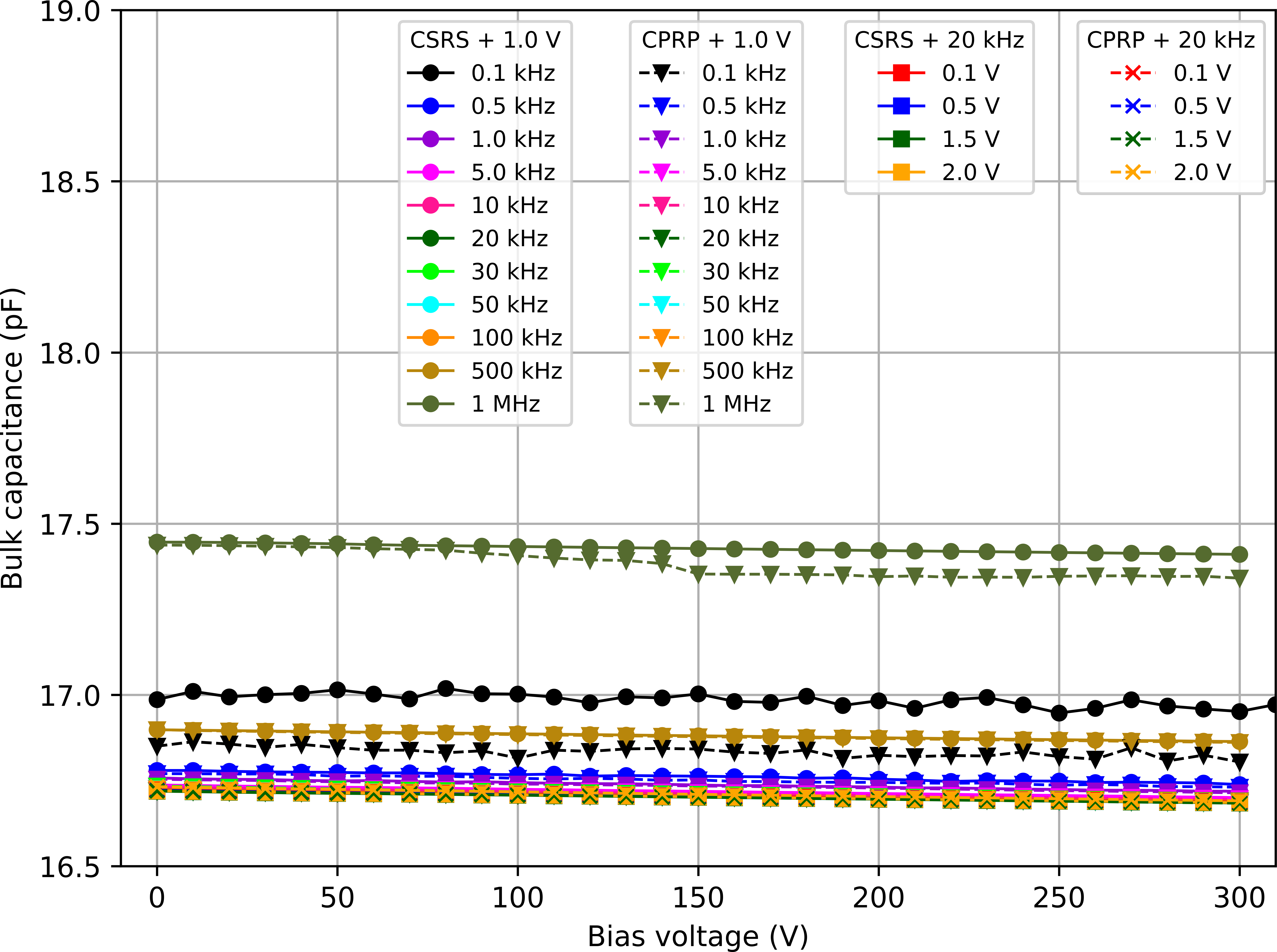}
  \includegraphics[width=0.49\textwidth]{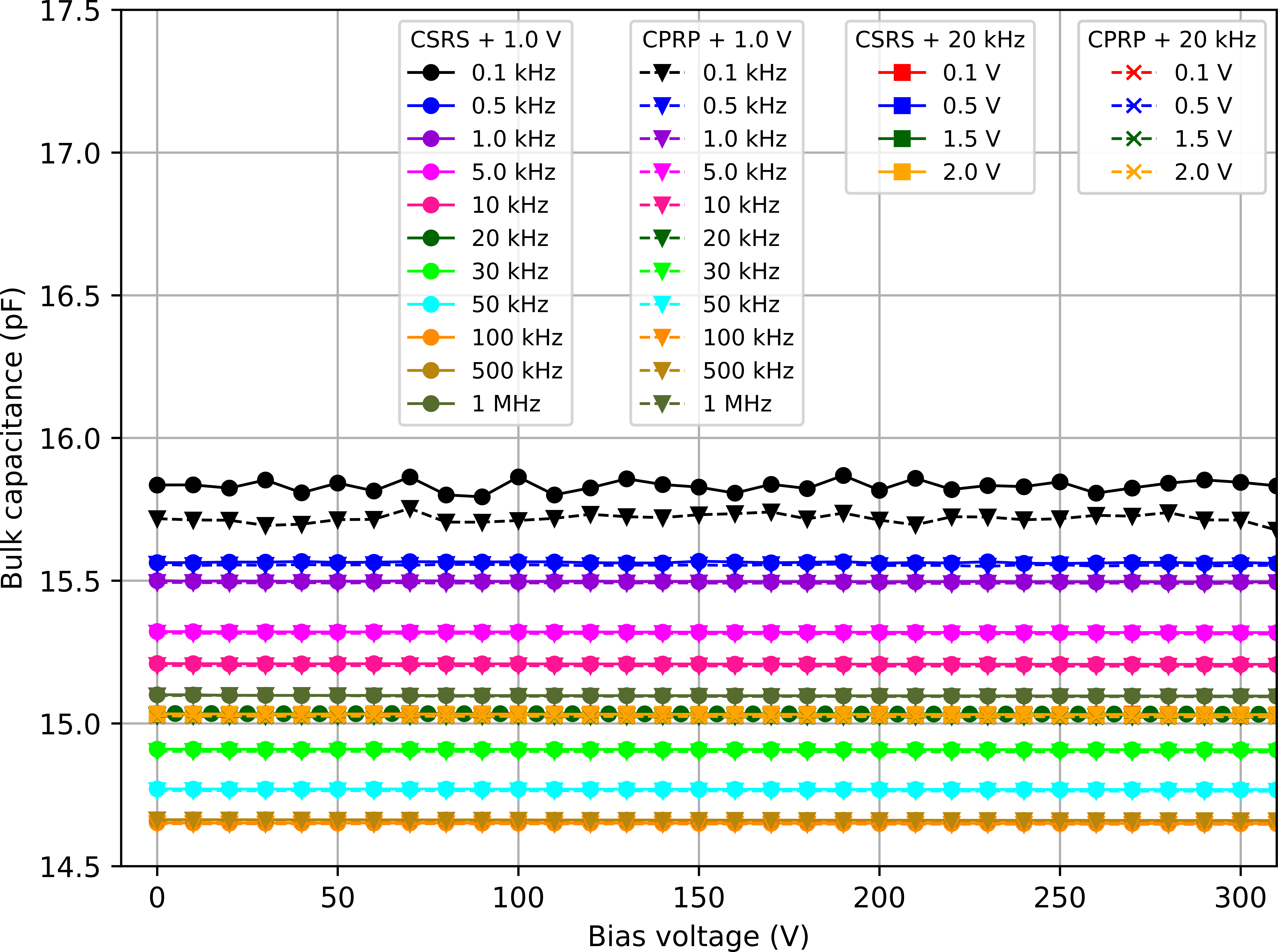}\hfill
  \includegraphics[width=0.49\textwidth]{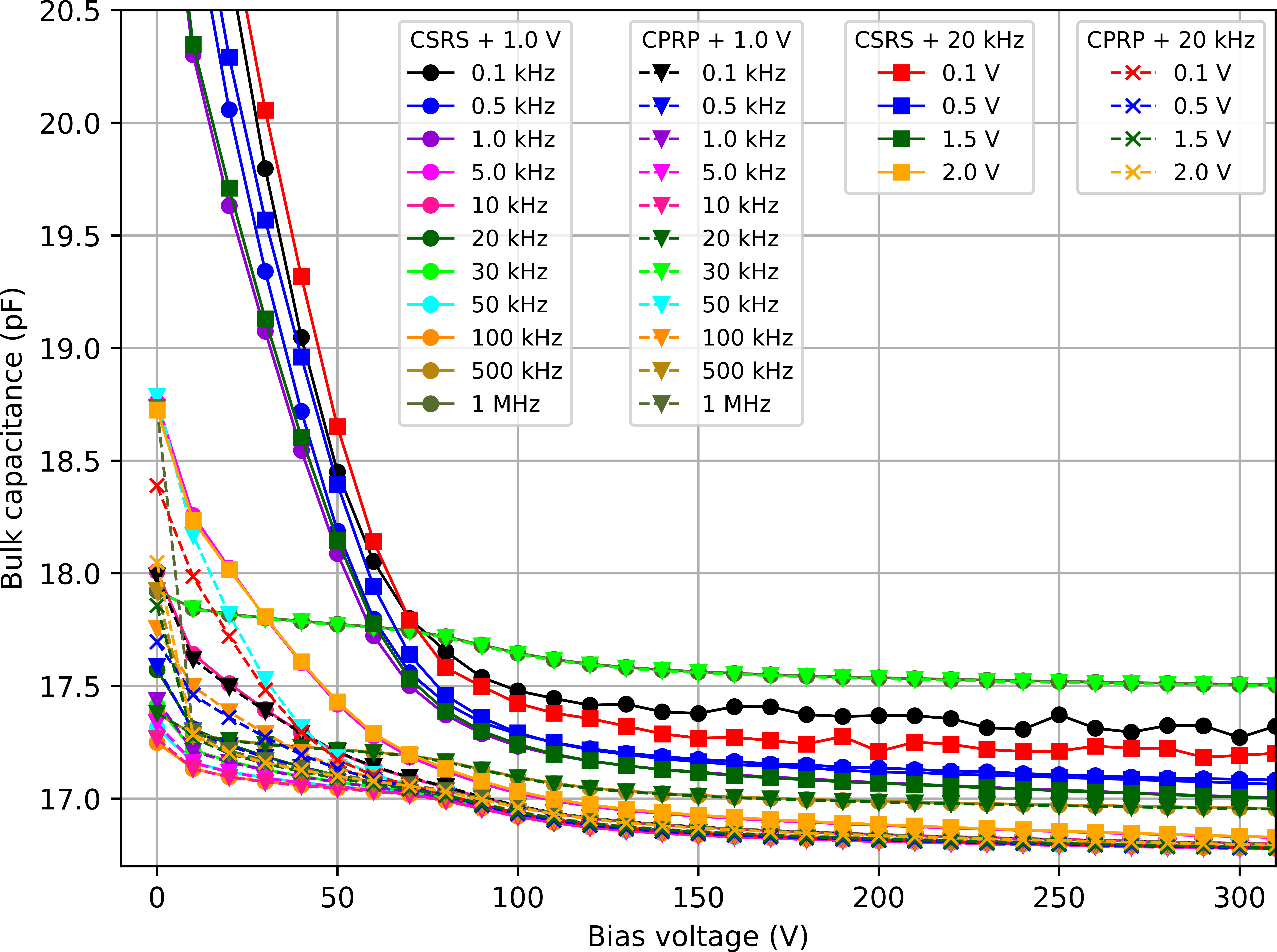}
  \caption{Bulk capacitance of 4H-SiC PN diodes measured as a function of reverse bias voltage. The influence of the LCR measurement configuration (CSRS and CPRP), as well as the frequency and amplitude of the sinusoidal AC test signal, is shown. The CV characteristics are presented for an unirradiated PN diode (a), a proton-irradiated PN diode exposed to $1.0 \times 10^{15}$~p/cm$^{2}$ (b), a neutron-irradiated PN diode exposed to $2.3 \times 10^{17}$~n$_{\mathrm{eq}}$/cm$^{2}$ (c), and a diode irradiated by $^{60}$Co gamma rays to 300~kGy (d). The thickness of the epitaxial layer is 50~$\mu$m for all displayed samples.}
  \label{fig:lcr_parameters}
\end{figure*}

Bulk capacitance was measured using an HP/Agilent 4284A Precision LCR meter. All CV measurements presented in this work were performed using the series measurement configuration (CSRS). The sinusoidal AC test signal frequency was set to 20~kHz and its amplitude to 1.0~V. The choice of measurement configuration, frequency, and signal amplitude was verified by systematic tests comparing CSRS and parallel (CPRP) configurations, as well as multiple frequencies and amplitudes. As shown in Fig.~\ref{fig:lcr_parameters}, no significant dependence of the measured bulk capacitance on these parameters was observed for unirradiated devices or for samples irradiated by protons and neutrons. For samples irradiated by $^{60}$Co gamma rays, a more pronounced dependence of bulk capacitance on the measurement configuration and test signal parameters is observed for bias voltages lower than \mbox{100 V}. The CSRS configuration with a test signal frequency of 20~kHz and amplitude of 1.0~V was selected for all CV measurements to ensure consistency across all sample types and irradiation conditions.

\section{Electrical characteristics of unirradiated 4H-SiC samples}
\label{results-unirradiated}
\noindent {The IV and CV characteristics were measured for 10 devices of each type: PN, PN-GRILL, LGAD1, LGAD1-GRILL, LGAD2, and LGAD2-GRILL, all diced from wafer W1. The label GRILL denotes the grill top metallization.}

\begin{figure*}[t]
  \centering
  \includegraphics[width=0.49\textwidth]{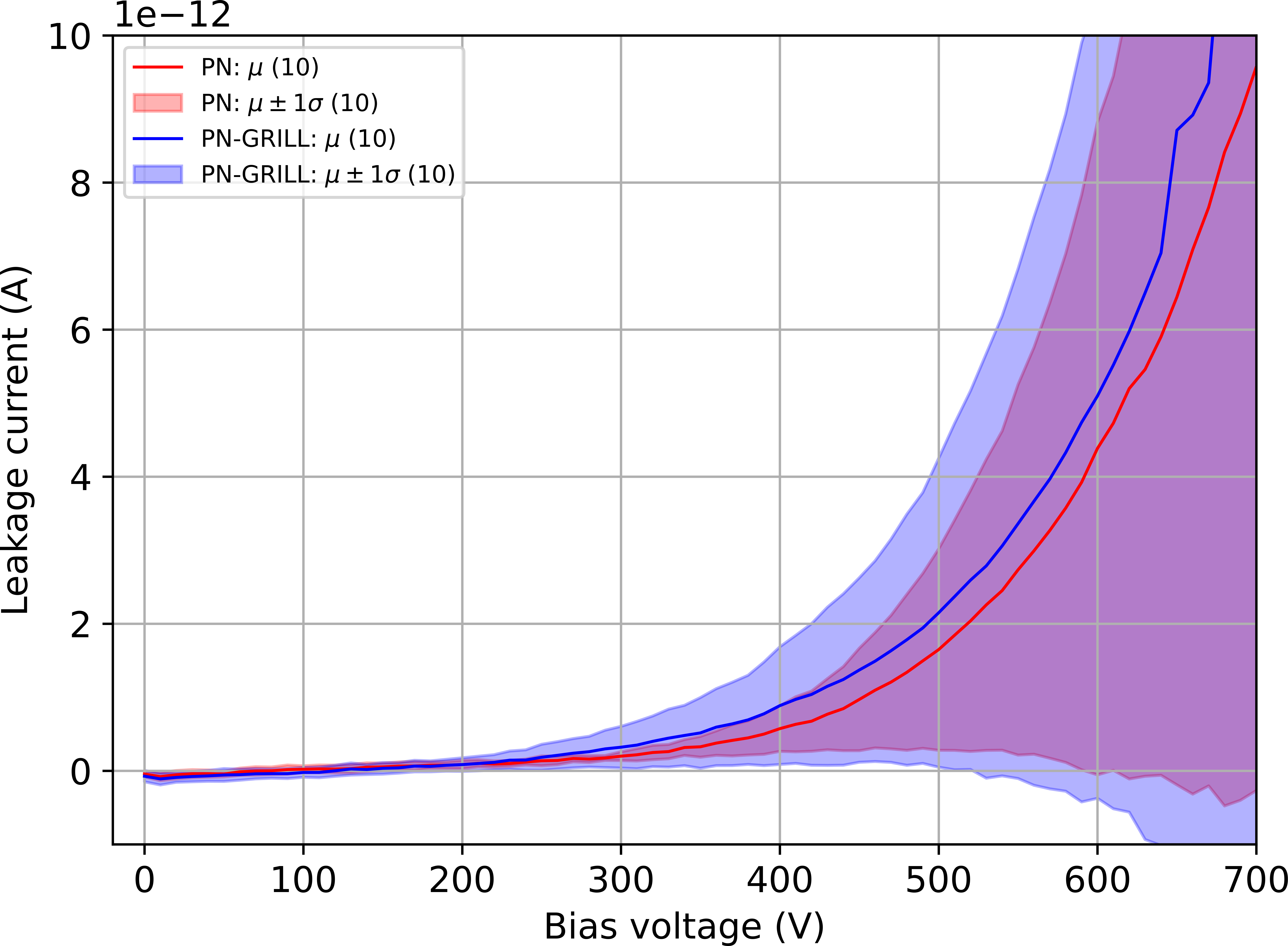} \hfill
  \includegraphics[width=0.49\textwidth]{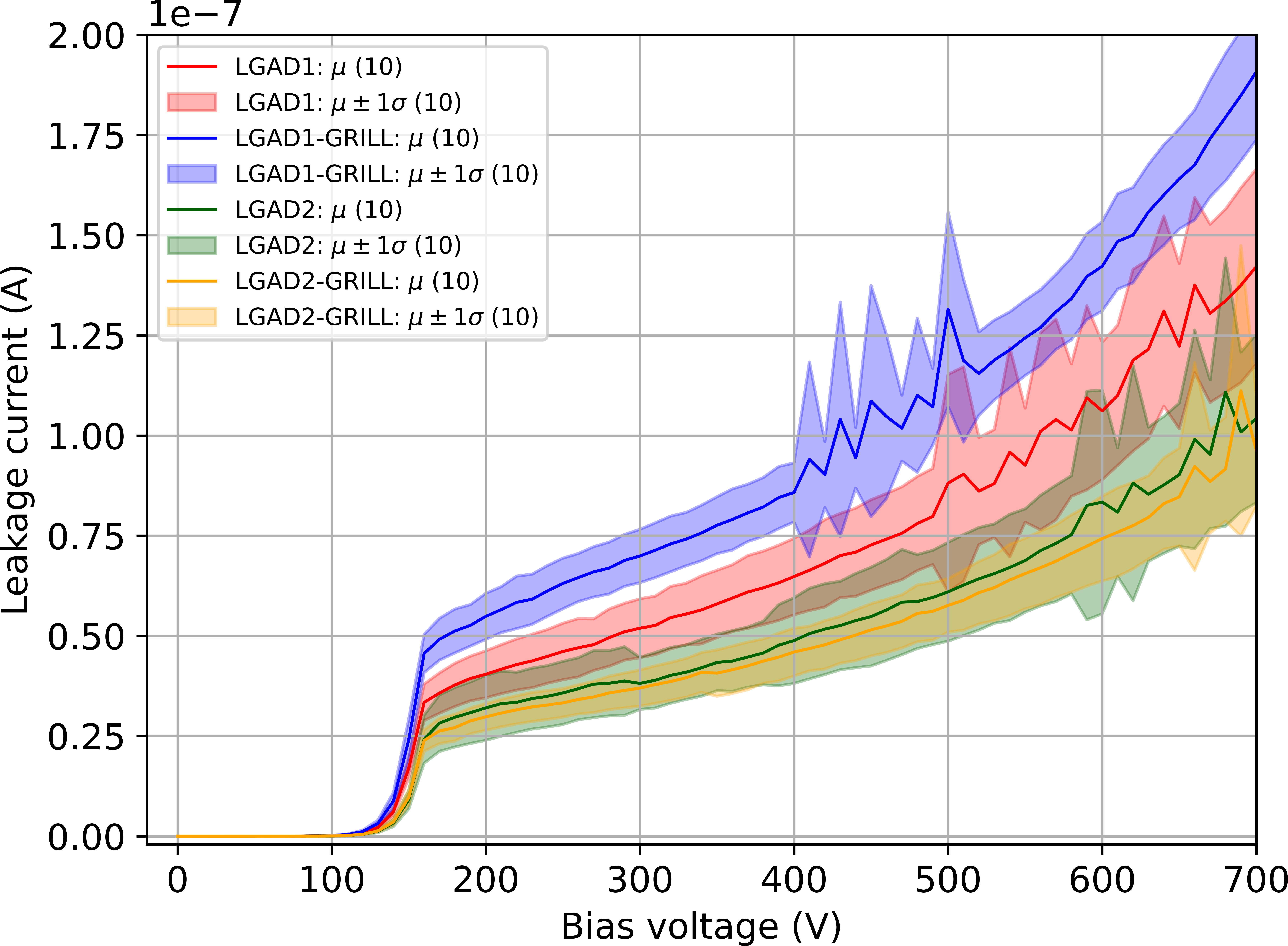}
  \caption{Leakage current as a function of applied bias voltage for unirradiated PN and LGAD devices. Solid lines represent the mean values calculated from the measured samples of each type (the number of measured samples of a given type is indicated in parentheses in the legend). The shaded regions correspond to the $\pm1\sigma$ intervals around the mean values.}
  \label{fig:iv_unirrad}
\end{figure*}

The IV characteristics measured for reverse bias voltages up to 700 V are shown in Fig.~\ref{fig:iv_unirrad}. For unirradiated PN diodes, the reverse leakage current reaches approximately 10~pA at 700~V. In contrast, LGAD devices exhibit reverse leakage currents exceeding 100~nA due to the presence of the internal multiplication layer. The forward characteristics are not shown, as the current compliance of 1~$\mu$A is reached for forward bias voltages between 1 and 2~V for all device types.

\begin{figure*}[t]
  \centering
  \includegraphics[width=0.325\textwidth]{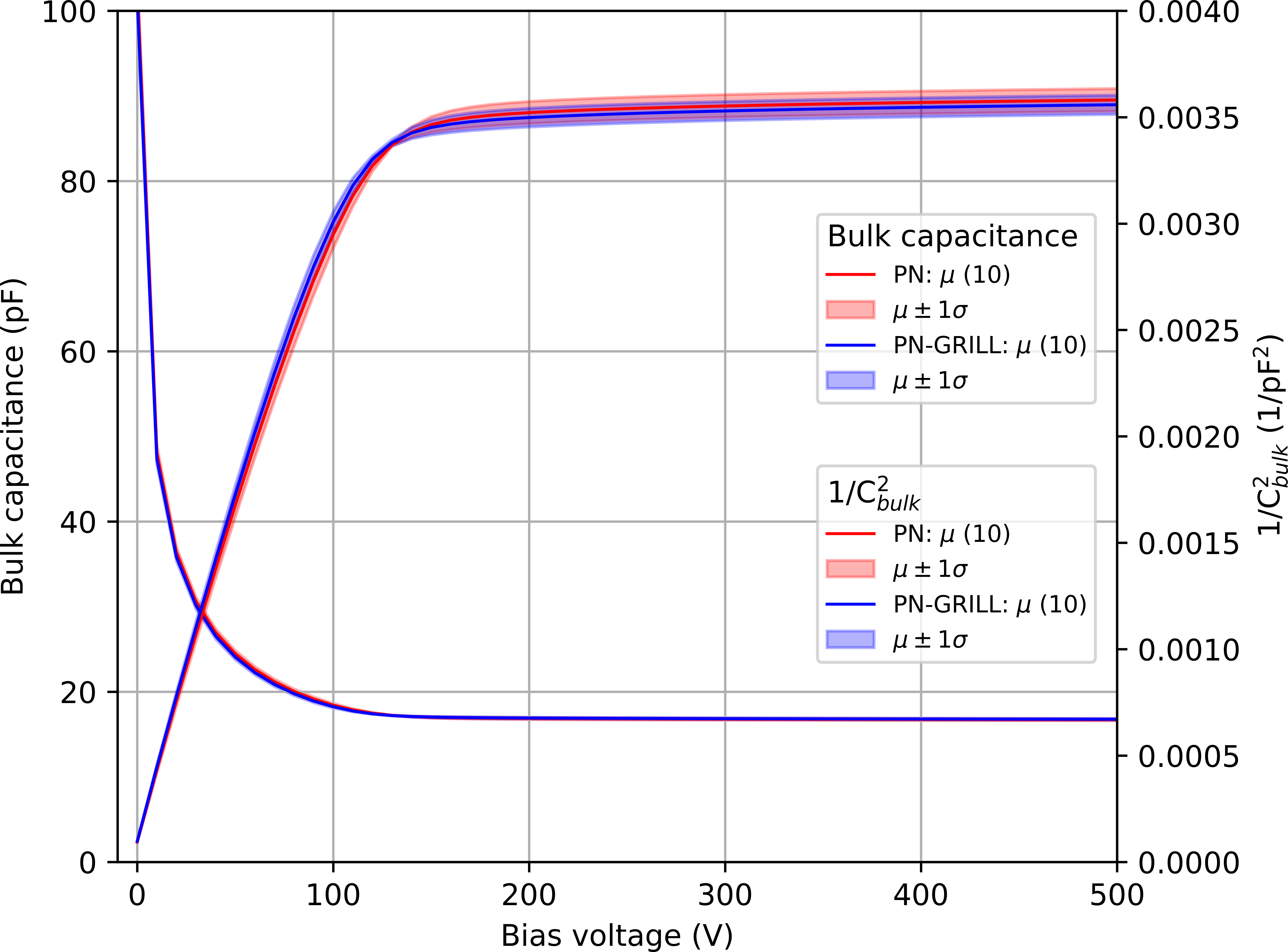} \hfill
  \includegraphics[width=0.325\textwidth]{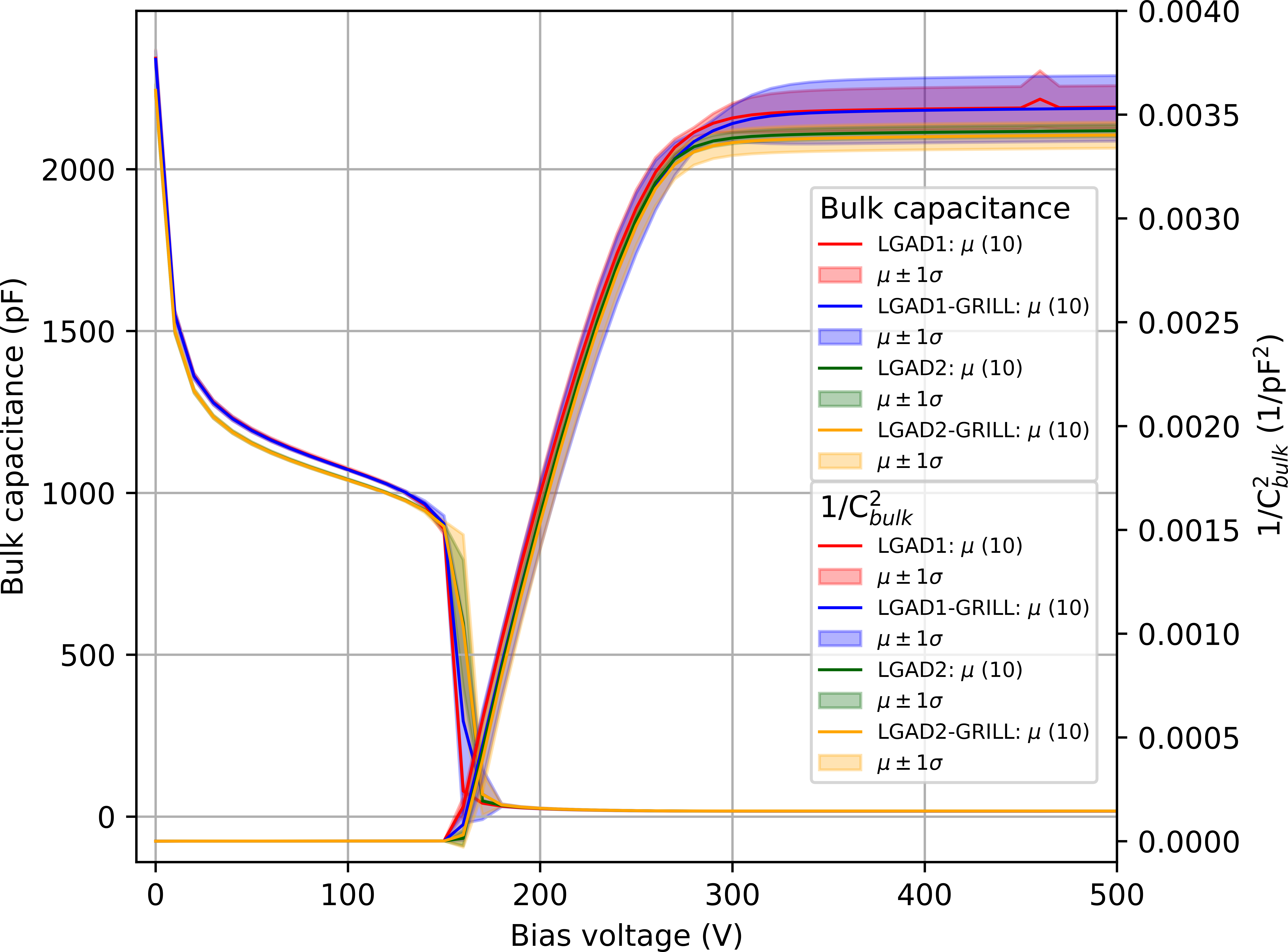} \hfill
  \includegraphics[width=0.325\textwidth]{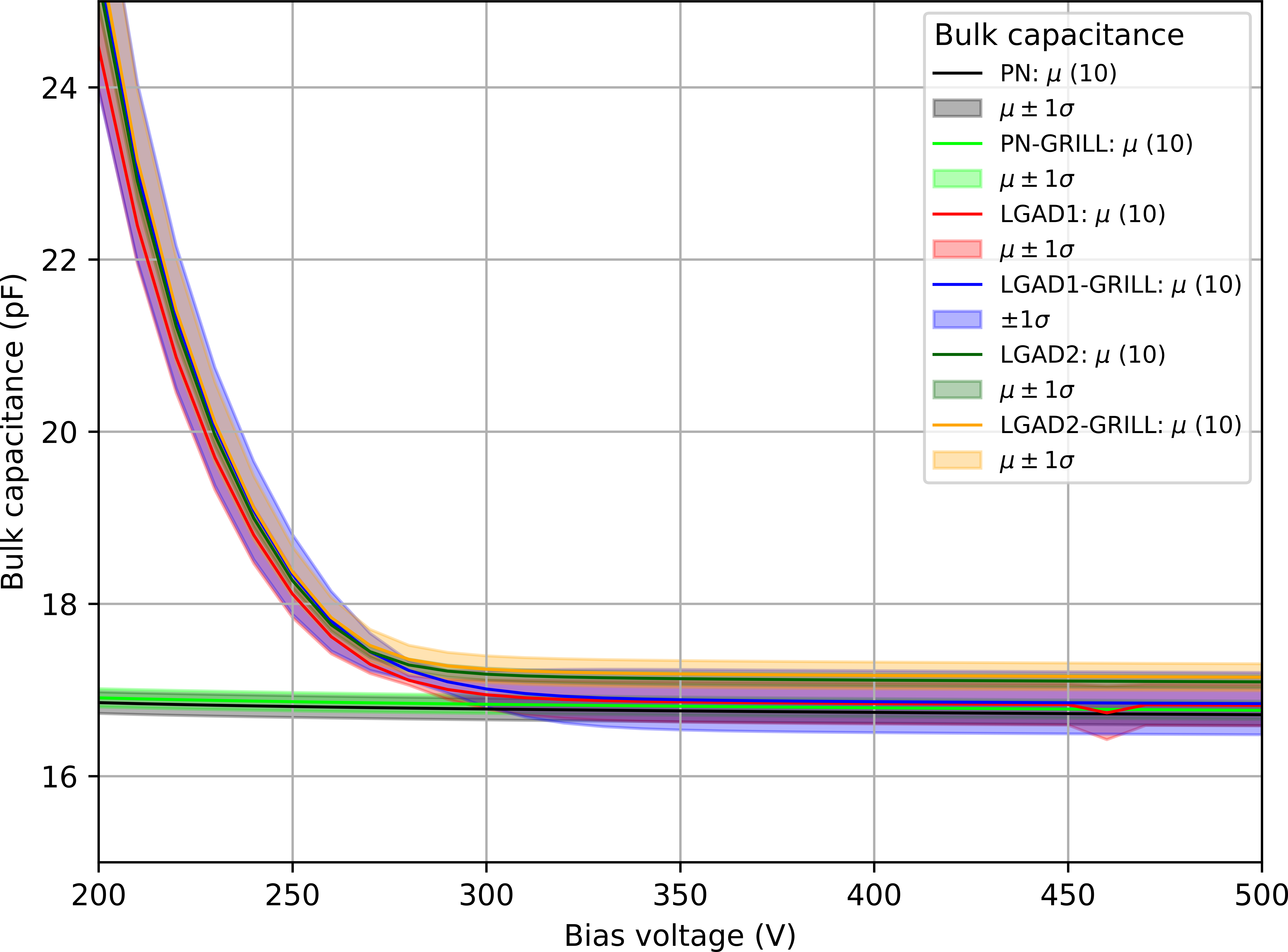} \hfill
  \caption{Bulk capacitance $C_{\mathrm{bulk}}$ and its inverse squared value $1/C_{\mathrm{bulk}}^{2}$ as a function of reverse bias voltage for unirradiated PN (left panel) and LGAD (middle panel) devices. An absolute value of the bulk capacitance of PN diodes and LGAD devices after their full depletion is shown in details (right panel). Solid lines represent mean values, and shaded regions denote $\pm1\sigma$ intervals, as defined in Fig.~\ref{fig:iv_unirrad}.}
  \label{fig:cv_unirrad}
\end{figure*}

The dependence of the bulk capacitance and its inverse squared value $1/C_{\mathrm{bulk}}^{2}$ on reverse bias voltage is presented in Fig.~\ref{fig:cv_unirrad} for voltages up to 500~V, well above the full depletion voltage. The spread between individual devices is minimal for all sample types, as indicated by the narrow $\pm1\sigma$ intervals. The PN diodes reach full depletion at approximately 120~V, whereas the full depletion voltage of the LGAD devices is close to 300~V. The slower decrease of capacitance observed between 20 and 160~V for LGAD devices corresponds to the progressive depletion of the highly doped multiplication layer.

After full depletion, the bulk capacitance reaches approximately 17~pF for both PN and LGAD devices. Using the standard relation between depletion capacitance, active area, and effective depletion width~\cite{lutz1999}, the fully depleted PN diode corresponds to an active thickness of approximately 46~$\mu$m and an effective doping concentration of approximately $6.2 \times 10^{13}$~cm$^{-3}$. These values are consistent with the nominal design parameters of the epitaxial layer described in Section~\ref{design}.  

\section{Electrical characteristics of irradiated 4H-SiC samples}
\noindent While the IV and CV characteristics of unirradiated 4H-SiC PN and LGAD devices are consistent with their design parameters, their behavior after exposure to different types of irradiation is critical for assessing their suitability for operation in radiation environments. Since no significant differences were observed between devices with full and grill top metallization prior to irradiation, this distinction is not considered in the following analysis of irradiated samples.

\subsection{Proton irradiated samples}
\begin{figure*}[t]
  \centering
  \includegraphics[width=0.49\textwidth]{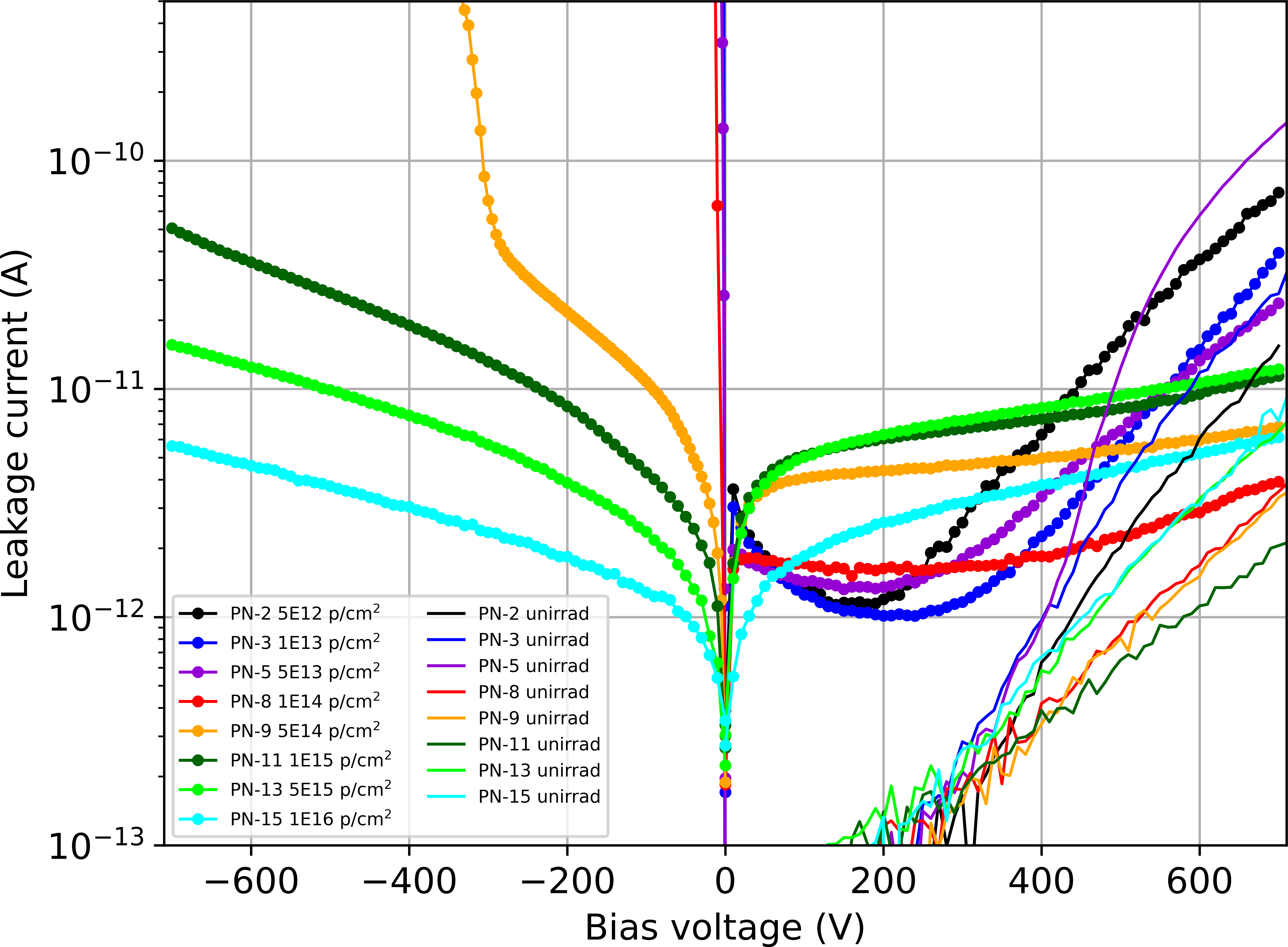} \hfill
  \includegraphics[width=0.49\textwidth]{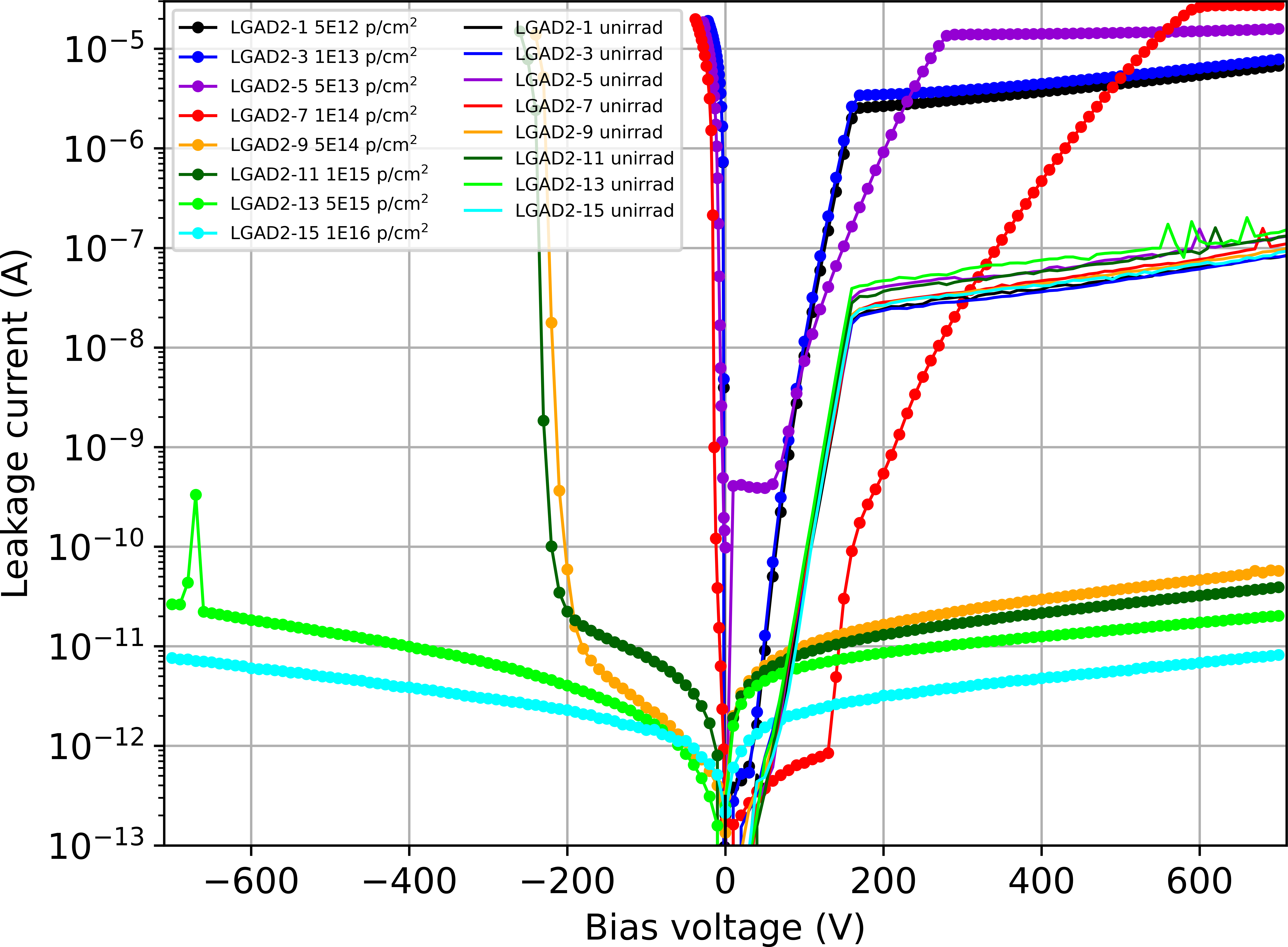}
  \caption{IV characteristics of PN (left) and LGAD2 (right) samples irradiated by \mbox{24 GeV/c} protons to eight different total proton fluences ranging from $5\times10^{12}$ to $1\times10^{16}\;\mathrm{p/cm^2}$. Solid lines with circle markers correspond to the IV curves measured after irradiation to the indicated fluences, while the solid lines represent the characteristics of the same samples measured prior to irradiation.}
  \label{fig:iv_protons}
\end{figure*}

\noindent The IV characteristics of PN and LGAD2 devices irradiated by 24~GeV/c protons at CERN IRRAD are shown in Fig.~\ref{fig:iv_protons} (LGAD1 devices exhibit equivalent behavior). The curves obtained after irradiation are compared with the pre-irradiation characteristics of the same devices.

For PN diodes, the reverse leakage current remains comparable to the unirradiated state at low proton fluences and high reverse bias voltages (above approximately 400~V), while an increase is observed at lower bias voltages. For total proton fluences exceeding $1 \times 10^{14}$~p/cm$^{2}$, the IV characteristics change significantly, and the reverse leakage current gradually decreases with increasing fluence. Even at room temperature, the reverse leakage currents remain substantially lower than those typically observed in silicon detectors irradiated to comparable fluences \cite{moll2018}. This behavior reflects the high displacement threshold energy of 4H-SiC and the reduced intrinsic carrier generation associated with its large bandgap \cite{rafi2023}.

In the forward bias region, the steep current increase observed for unirradiated and low-fluence samples progressively weakens with increasing proton fluence. For the highest fluences, the IV characteristics become nearly symmetric with respect to voltage polarity. This behavior is consistent with strong radiation-induced compensation of the initially N-type material by acceptor-like deep-level defects (such as Z$_{1/2}$ and EH$_{6/7}$ centers) \cite{kaneko2011}, leading to a substantial reduction of the effective free carrier concentration and an increase in resistivity. 

The LGAD devices exhibit a pronounced increase in reverse leakage current for low and intermediate proton fluences, reaching values of several tens of~$\mu$A. However, for fluences exceeding $1 \times 10^{14}$~p/cm$^{2}$, a significant reduction of the reverse leakage current is observed, spanning more than six orders of magnitude across the investigated fluence range. The forward characteristics follow a trend similar to that observed for PN diodes, with nearly symmetric behavior at the highest fluences.

\begin{figure*}[t]
  \centering
  \includegraphics[width=0.325\textwidth]{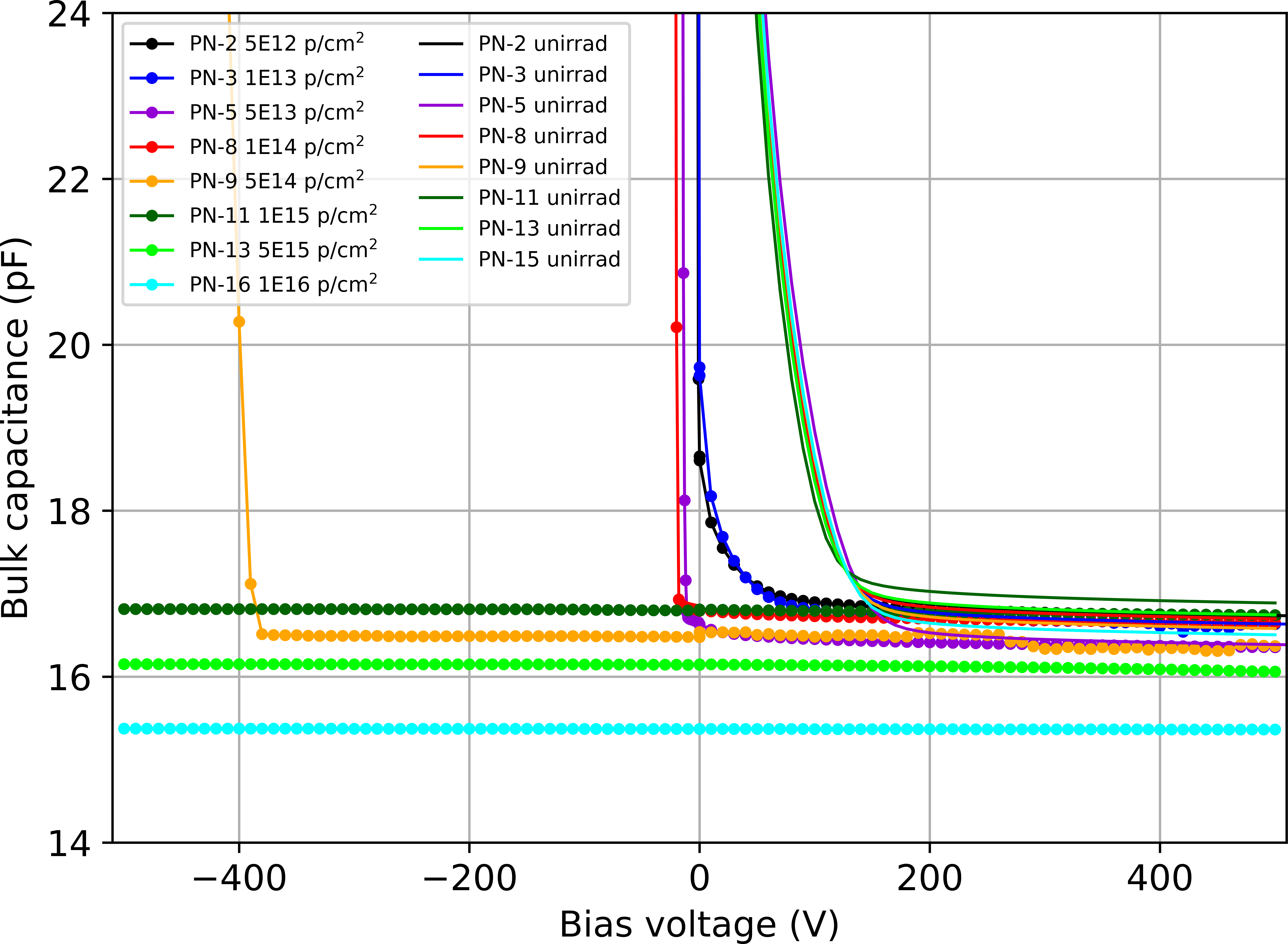} \hfill
  \includegraphics[width=0.325\textwidth]{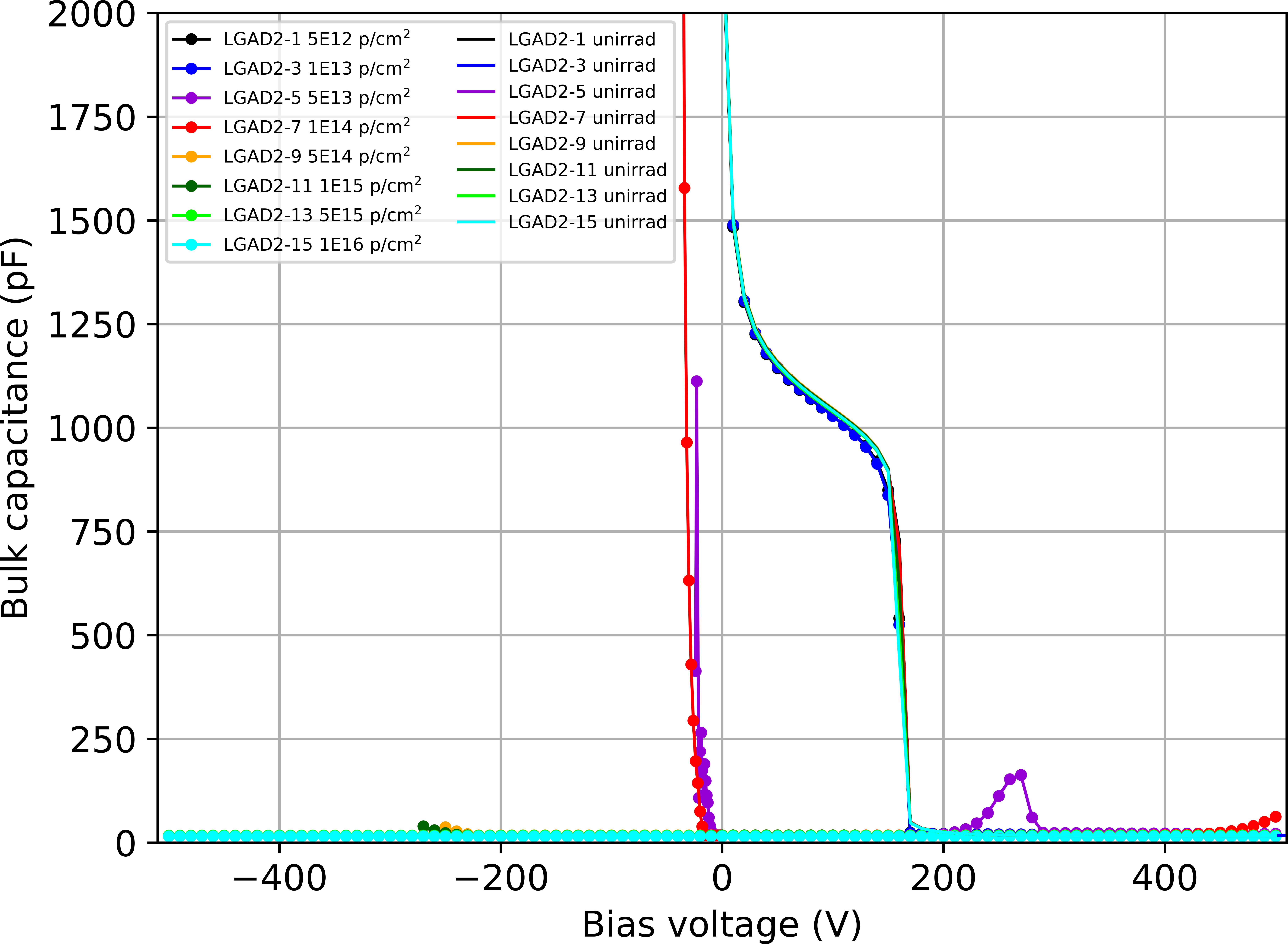} \hfill
  \includegraphics[width=0.325\textwidth]{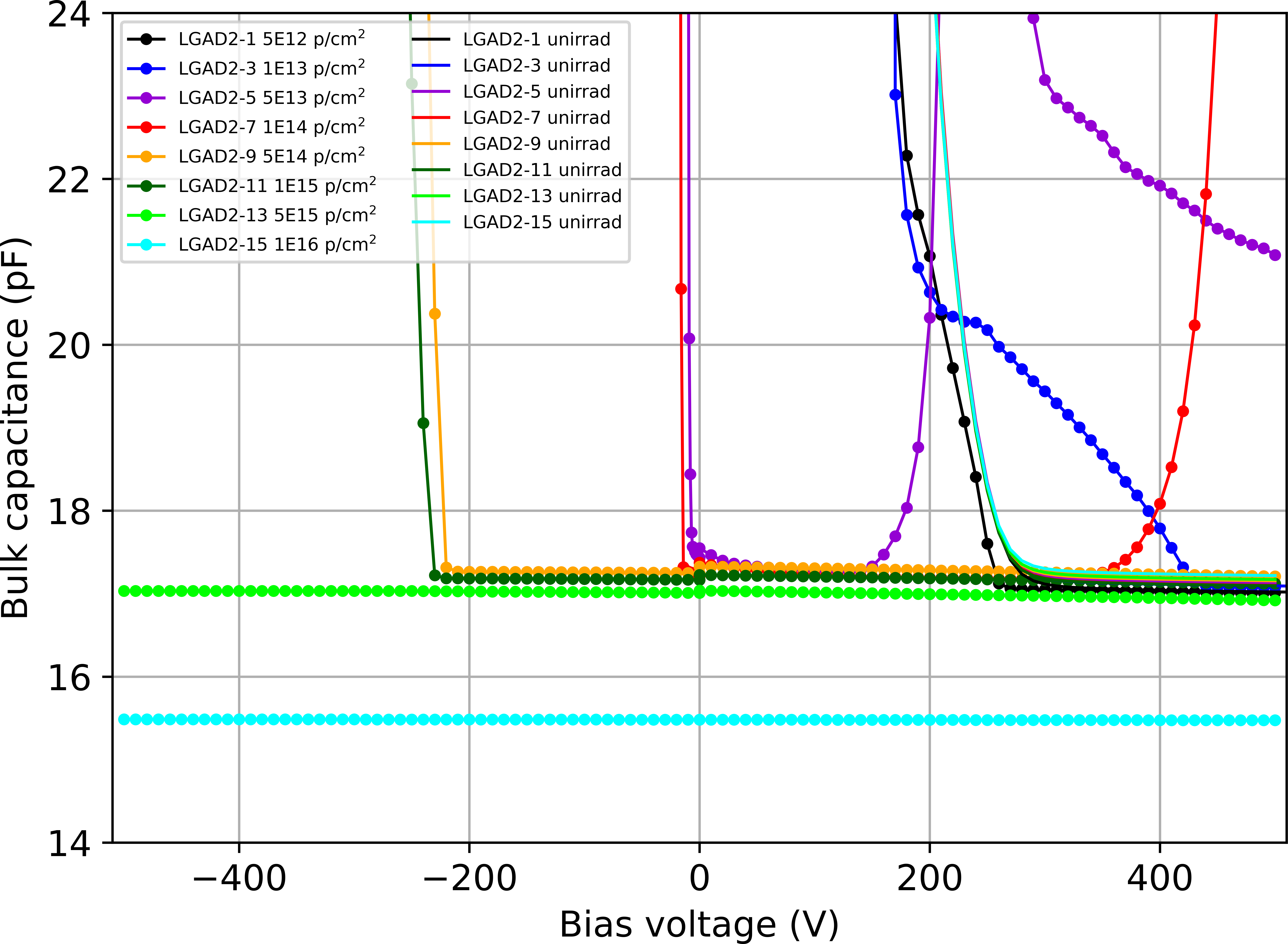}
  \caption{Bulk capacitance of PN (left) and LGAD2 (middle, right) samples irradiated by \mbox{24 GeV/c} protons to eight different total proton fluences ranging from $5\times10^{12}$ to $1\times10^{16}\;\mathrm{p/cm^2}$. The middle panel shows the full range of measured capacitance values to illustrate the overall trend, while the smaller y-axis range used in the left and right panels highlights the changes in the CV characteristics with increasing proton fluence. The notation is equivalent to that used in Fig.~\ref{fig:iv_protons}.}
  \label{fig:cv_protons}
\end{figure*}

The evolution of the bulk capacitance with bias voltage, shown in Fig.~\ref{fig:cv_protons}, supports the interpretation based on radiation-induced compensation. For low fluences, the CV characteristics retain the diode-like behavior observed prior to irradiation, although the capacitance decreases more steeply with reverse bias. Starting from fluences of approximately $5 \times 10^{13}$~p/cm$^{2}$, the bulk capacitance becomes nearly independent of bias voltage over a wide voltage range. The constant capacitance corresponds approximately to the geometrical capacitance of the fully depleted epitaxial layer with a thickness of 50~$\mu$m. 

For the highest investigated proton fluence of $1 \times 10^{16}$~p/cm$^{2}$, the extracted bulk capacitance is slightly reduced, reaching values around 15.4~pF after full depletion. This modest decrease is consistent with the strong radiation-induced compensation of the originally low-doped epitaxial layer, which leads to a nearly uniform electric field distribution across its thickness.

\subsection{Neutron irradiated samples}
\begin{figure*}[t] 
  \centering
  \includegraphics[width=0.325\textwidth]{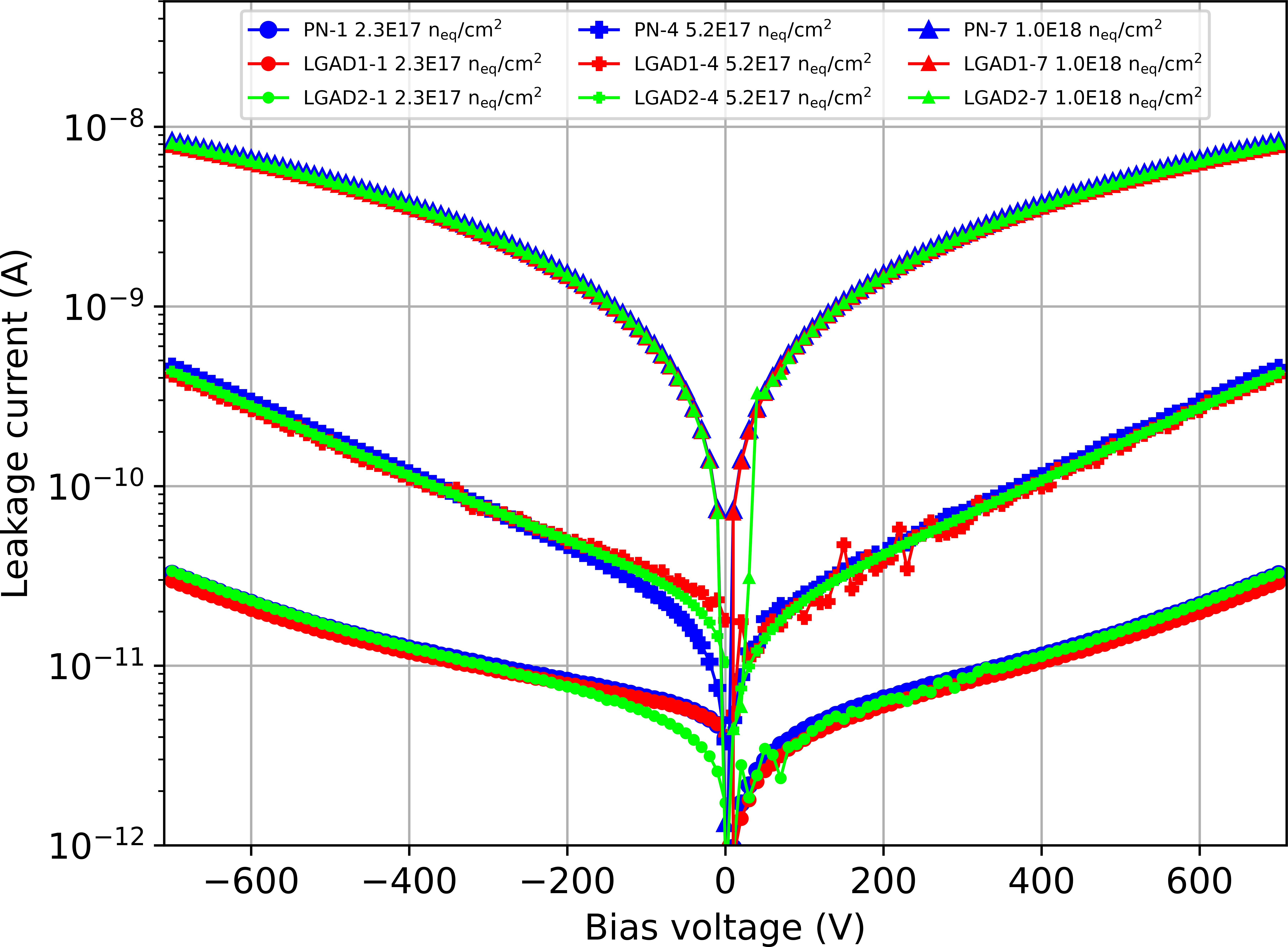} \hfill
  \includegraphics[width=0.325\textwidth]{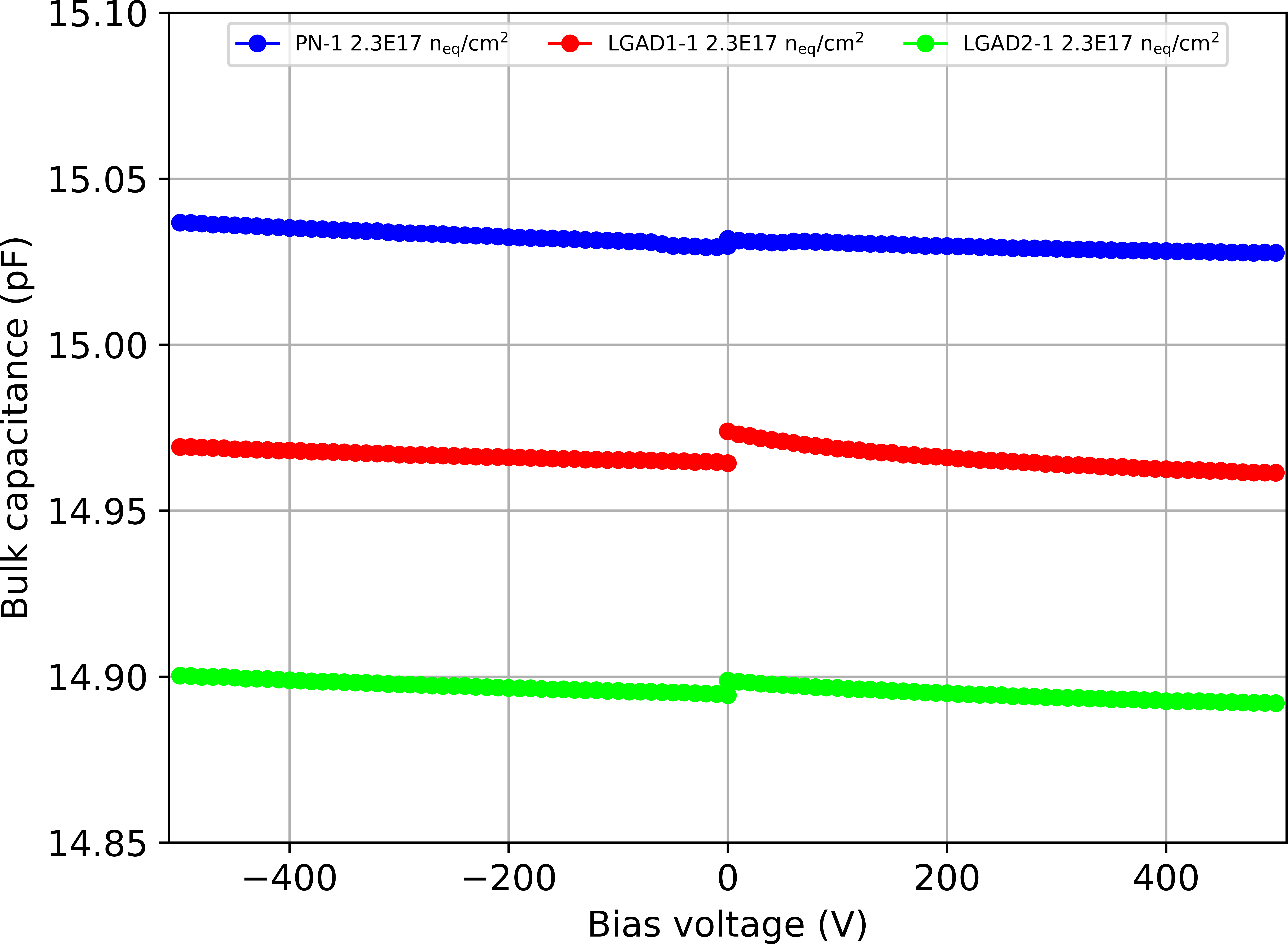} \hfill
  \includegraphics[width=0.325\textwidth]{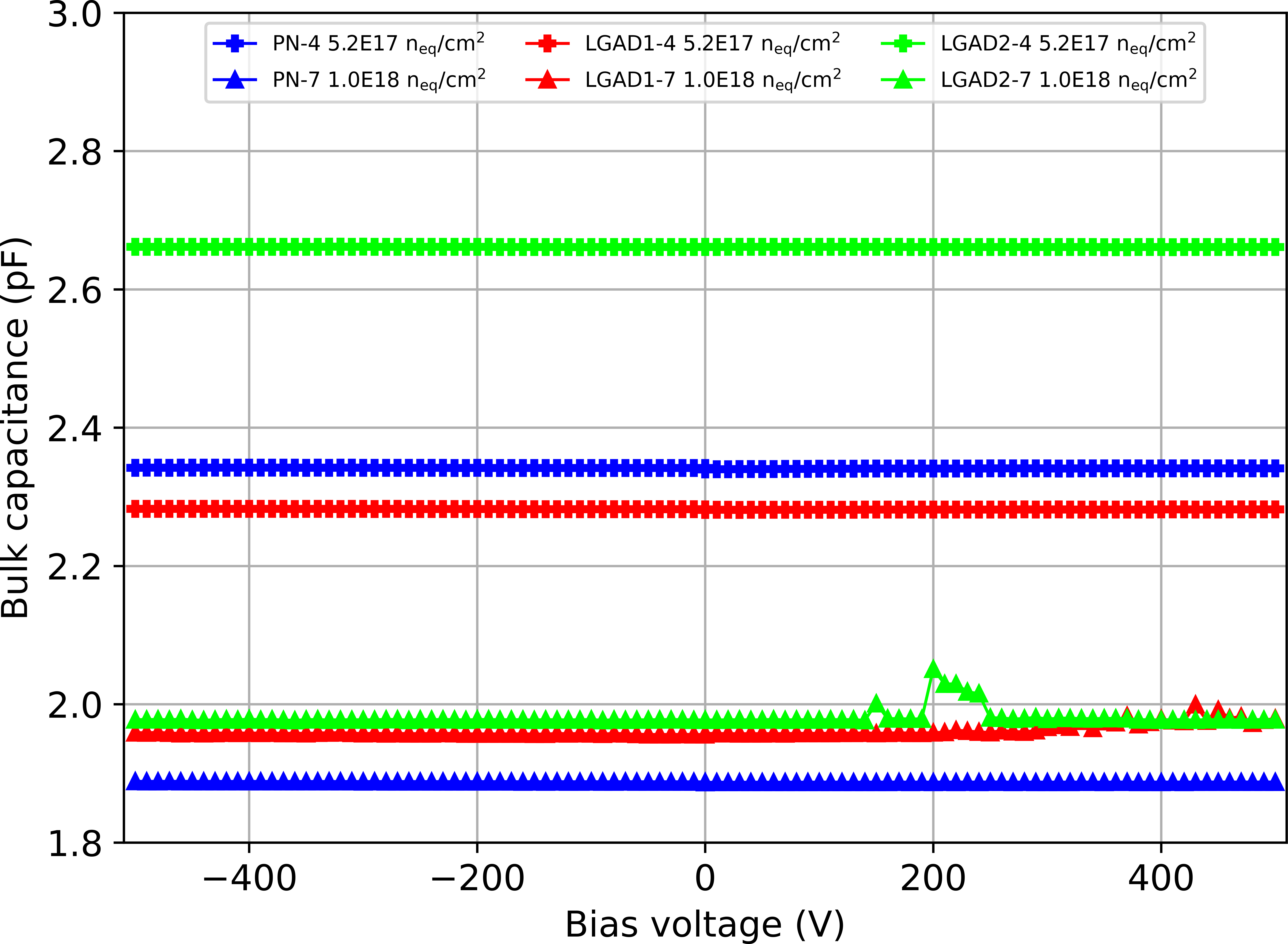}
  \caption{IV and CV characteristics of 4H-SiC PN and LGAD diodes irradiated by neutrons. The IV characteristics for all device types and neutron fluences are shown in the left panel. The middle panel presents the CV characteristics for PN, LGAD1, and LGAD2 samples irradiated to \mbox{$2.3\times10^{17}\;\mathrm{n_{eq}/cm^2}$}. The right panel shows the CV characteristics for the higher neutron fluences of \mbox{$5.2\times10^{17}$} and \mbox{$1.0\times10^{18}\;\mathrm{n_{eq}/cm^2}$}.}
  \label{fig:iv_cv_neutrons}
\end{figure*}

\noindent The IV and CV characteristics measured for 4H-SiC PN diodes and LGADs irradiated by reactor neutrons to extreme neutron fluences of \mbox{$2.3\times10^{17}$}, \mbox{$5.2\times10^{17}$}, and \mbox{$1.0\times10^{18}\;\mathrm{n_{eq}/cm^2}$} are shown in Fig.~\ref{fig:iv_cv_neutrons}. The leakage current measured for PN, LGAD1, and LGAD2 samples is practically identical and reaches similar absolute values in both reverse and forward bias directions. For the fluence of \mbox{$2.3\times10^{17}\;\mathrm{n_{eq}/cm^2}$}, the leakage current is approximately $30\;\mathrm{pA}$ at $\pm700\;\mathrm{V}$, slightly above the currents observed for the highest proton fluences. The increase of the leakage current becomes more pronounced for higher neutron fluences, reaching approximately $0.5\;\mathrm{nA}$ for \mbox{$5.2\times10^{17}\;\mathrm{n_{eq}/cm^2}$} and almost $10\;\mathrm{nA}$ for \mbox{$1.0\times10^{18}\;\mathrm{n_{eq}/cm^2}$}.

The CV characteristics measured for the neutron fluence of \mbox{$2.3\times10^{17}\;\mathrm{n_{eq}/cm^2}$} remain broadly consistent with the behavior observed for proton-irradiated samples. The bulk capacitance after depletion is close to the geometrical capacitance of the epitaxial layer, reaching values around 15~pF. A qualitatively different behavior is observed for the highest neutron fluences of \mbox{$5.2\times10^{17}$} and \mbox{$1.0\times10^{18}\;\mathrm{n_{eq}/cm^2}$}. In these cases, the extracted bulk capacitance decreases to approximately \mbox{2.2~pF}, corresponding geometrically to a depleted thickness of about \mbox{350~$\mu$m}. Considering that the epitaxial layer thickness is only $50~\mu$m, this result indicates that the depletion region extends far beyond the epitaxial layer and penetrates deeply into the originally highly doped substrate.

The capacitance-derived depletion width corresponds to an effective doping concentration on the order of $10^{12}$--$10^{13}\;\mathrm{cm^{-3}}$~\cite{lutz1999,moll2018}, indicating very strong modification of the effective space charge in the irradiated material. While the nominal donor concentration of the epitaxial layer is approximately $5\times10^{13}\;\mathrm{cm^{-3}}$, the much larger depletion width extracted from the capacitance measurements implies that the effective space charge density becomes very small also in regions extending well into the substrate, which originally had a donor concentration of the order of $10^{19}\;\mathrm{cm^{-3}}$. This observation suggests substantial radiation-induced compensation and defect-related modification of the space charge in heavily irradiated 4H-SiC~\cite{kaneko2011}, which can significantly alter the effective carrier concentration and electric-field distribution in the device.

The resulting enlargement of the effective depletion volume significantly increases the active material volume contributing to leakage current. However, the observed leakage current level cannot be explained by simple bulk thermal SRH generation~\cite{shockley1952} in the depletion region, given the extremely low intrinsic carrier concentration of 4H-SiC. This suggests that, in addition to the enlarged depletion volume, field-enhanced carrier generation and/or surface and edge-related leakage mechanisms also contribute to the measured current.

\subsection{Samples irradiated by $^{60}$Co gamma rays}
Irradiation of 4H-SiC material with gamma rays from a $^{60}$Co source, emitting photons with energies of \mbox{1.17 MeV} and \mbox{1.33 MeV}, causes radiation damage predominantly through Compton electrons generated with energies up to about \mbox{1 MeV}. These electrons can produce point-like displacement defects in the crystal lattice; however, their energy is generally insufficient to generate displacement cascades or cluster defects. At the same time, gamma irradiation is known to induce significant ionization damage, leading to the formation of charge traps in oxide layers, semiconductor–oxide interfaces, and passivation layers.

\begin{figure*}[t] 
  \centering
  \includegraphics[width=0.49\textwidth]{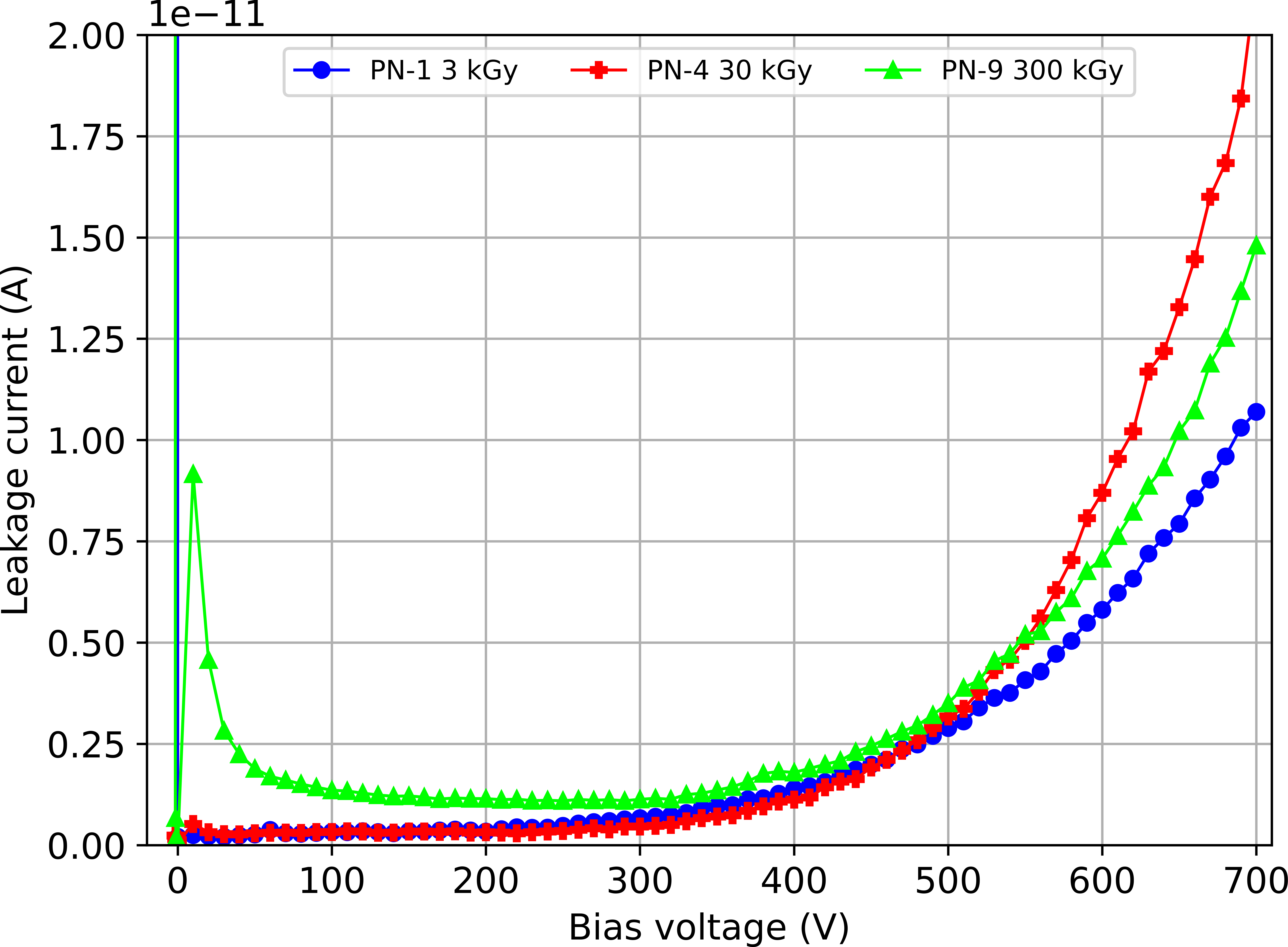} \hfill
  \includegraphics[width=0.49\textwidth]{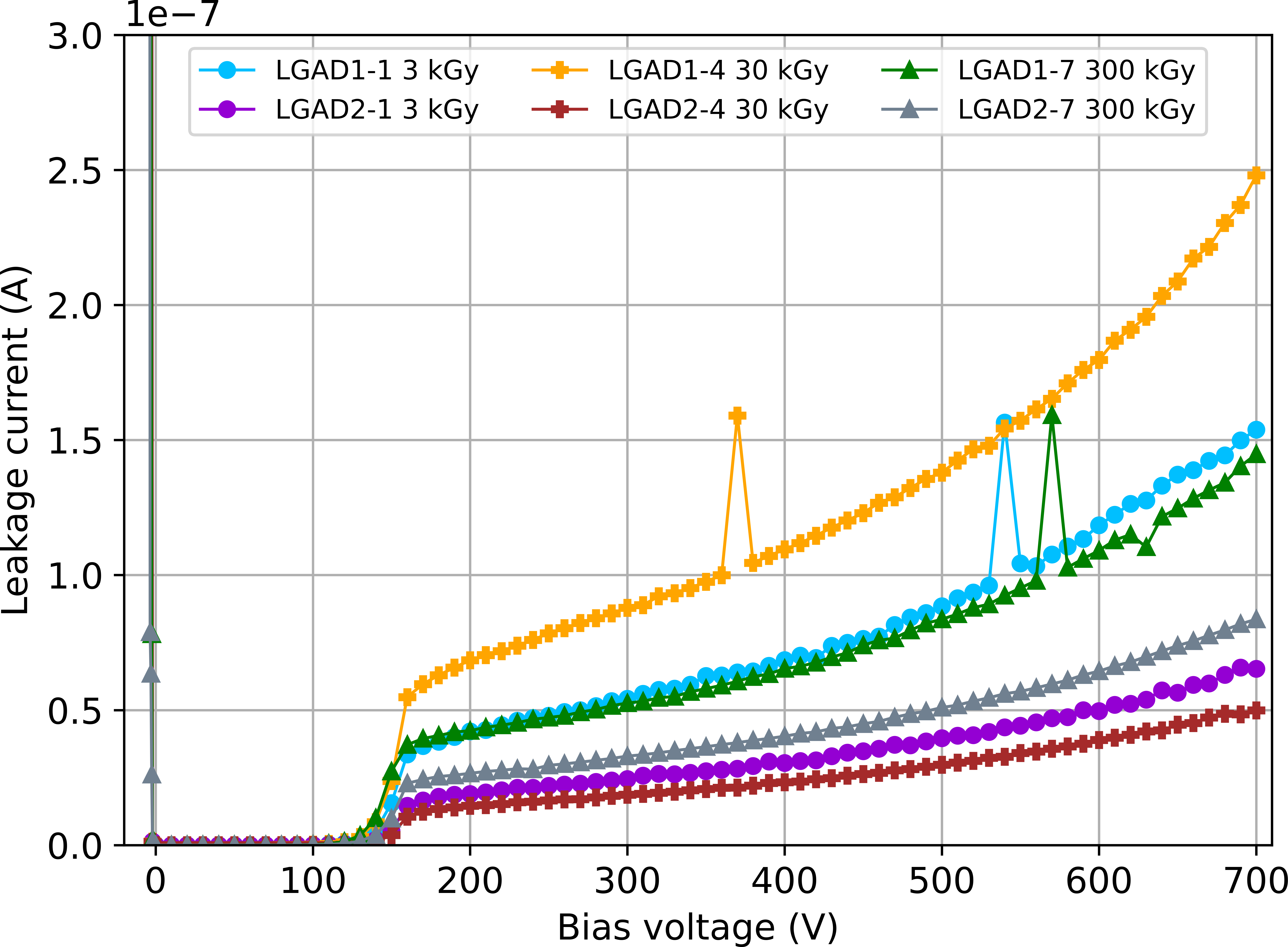}
  \vspace{0.4em}
  \includegraphics[width=0.49\textwidth]{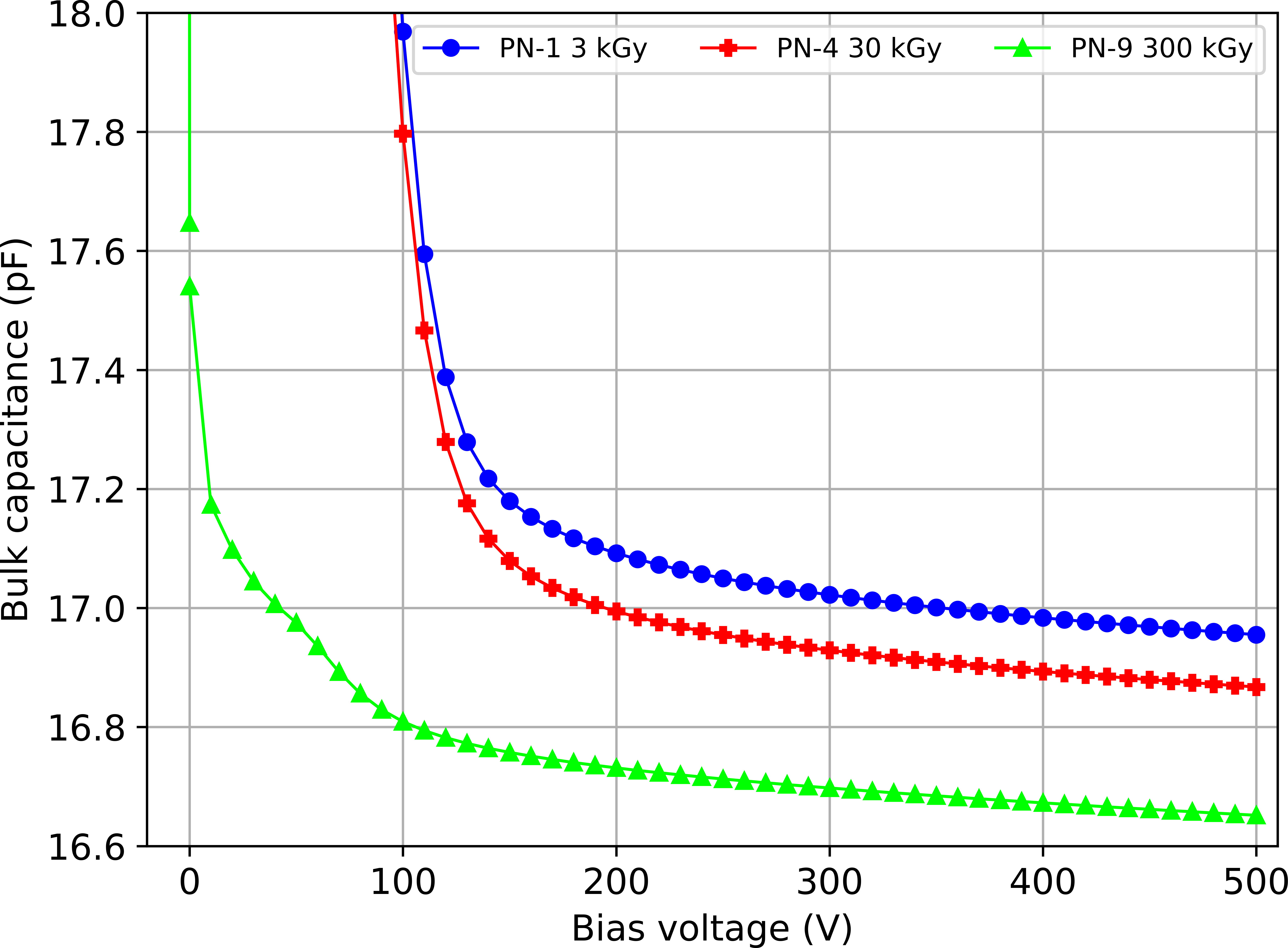} \hfil
  \includegraphics[width=0.49\textwidth]{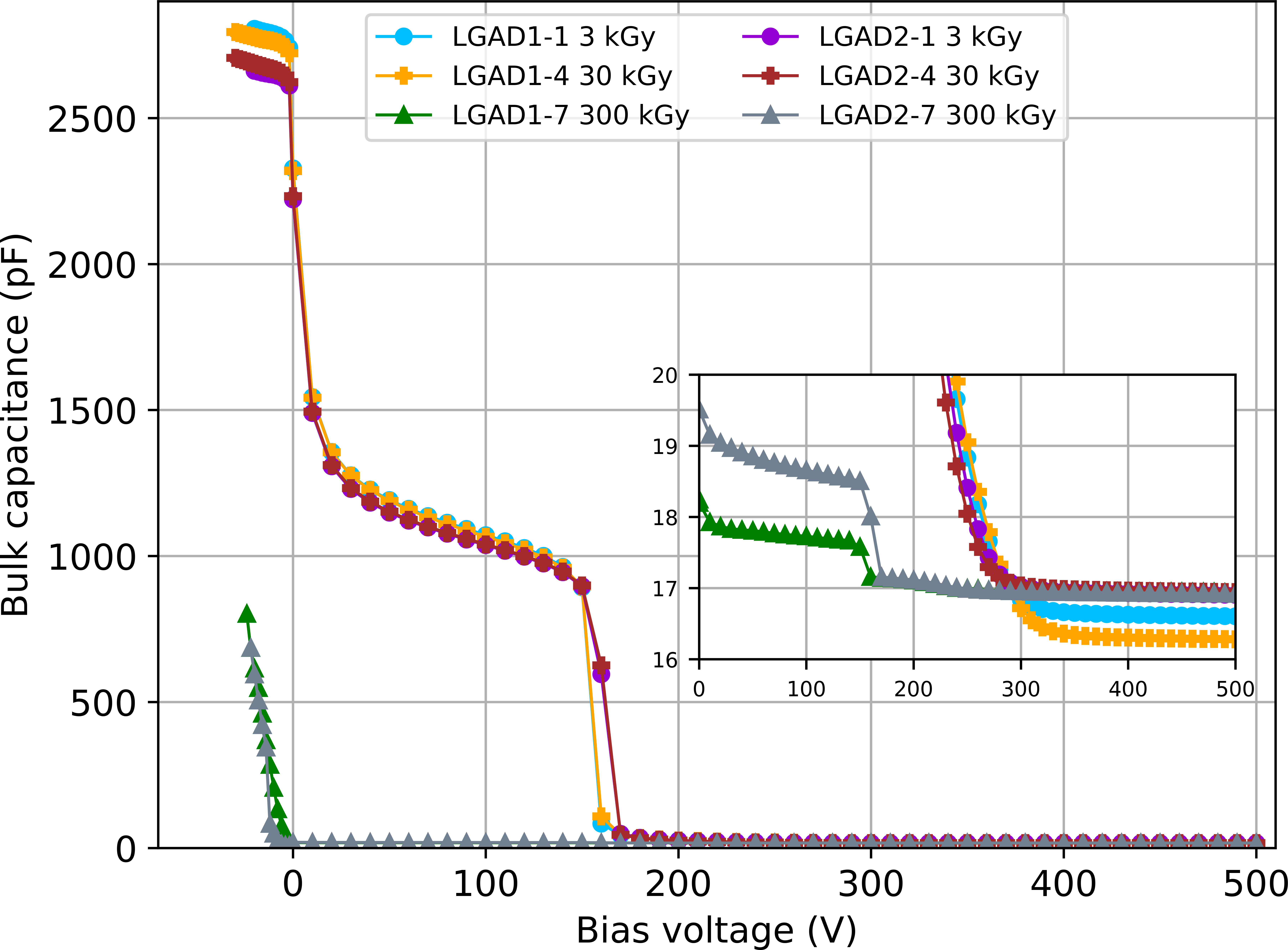}
  \caption{IV and CV characteristics of PN diodes (top and bottom left) and LGAD sensors (top and bottom right) irradiated by $^{60}$Co gamma rays to TIDs of 3, 30, and \mbox{300 kGy}.}
  \label{fig:iv_cv_co60}
\end{figure*}

The IV and CV characteristics of PN, LGAD1, and LGAD2 samples after irradiation by $^{60}$Co gamma rays to total ionizing doses (TIDs) of 3, 30, and \mbox{300 kGy} are shown in Fig.~\ref{fig:iv_cv_co60}. The reverse leakage current of both PN and LGAD devices remains statistically comparable to the behavior observed before irradiation.  In the forward-bias direction (negative voltage polarity), most samples reach the set current compliance limit of $1\;\mu\mathrm{A}$ immediately after the bias is applied. However, several devices irradiated to 30 and 300 kGy exhibit an anomalous behavior in which the forward current remains near the measurement floor up to bias voltages of several tens of volts, followed by an abrupt increase to the compliance limit. Such threshold-like conduction is inconsistent with bulk displacement damage, which would be expected to produce gradual and monotonic changes of the forward characteristics with increasing dose. Instead, this behavior suggests radiation-induced interface or oxide-related effects that may lead to local electric-field redistribution or to the formation of localized conduction paths in the device periphery.   

The effect of $^{60}$Co irradiation on the measured bulk capacitance is more pronounced. For PN diodes irradiated to 3 and 30 kGy, the CV characteristics exhibit the typical diode-like behavior observed for unirradiated samples. A clear modification appears only for the highest investigated dose of 300 kGy, where the decrease of the bulk capacitance with increasing reverse bias becomes significantly steeper. After reaching the full depletion voltage, the capacitance stabilizes at approximately $16.8\;\mathrm{pF}$, corresponding to the geometrical capacitance of the fully depleted $50\;\mu$m thick epitaxial layer.

For LGAD devices, the CV characteristics remain similar to those of unirradiated samples for doses of 3 and \mbox{30 kGy}, showing the characteristic depletion of the multiplication layer followed by depletion of the epitaxial region. For the highest dose of 300 kGy, however, the capacitance becomes nearly constant over the entire range of reverse bias voltages, reaching values around $17\;\mathrm{pF}$. 
This behavior indicates a substantial reduction of the effective doping concentration in both the epitaxial region and the internal multiplication layer. 

\section{Summary and conclusions}
\label{conclusions}
\noindent The electrical characteristics of 4H-SiC PN, LGAD1, and LGAD2 devices fabricated on wafer W1 were investigated before and after irradiation by \mbox{24 GeV/c} protons, reactor neutrons, and $^{60}$Co gamma rays. {The pre-irradiation IV and CV characteristics are consistent with the device design: the reverse leakage current reaches approximately \mbox{10 pA} for PN diodes and \mbox{100 nA} for LGAD devices}, while the bulk capacitance after full depletion is close to \mbox{17 pF}, corresponding to the geometrical capacitance of the \mbox{50 $\mu$m} thick epitaxial layer. The full depletion voltage is approximately \mbox{120 V} for PN diodes and \mbox{300 V} for LGAD devices, with the latter showing the expected two-step depletion associated with the multiplication layer and the epitaxial region.

Irradiation by high-energy protons leads to pronounced changes in both IV and CV characteristics. For proton fluences exceeding approximately \mbox{$1\times10^{14}\;\mathrm{p/cm^2}$}, the reverse leakage current decreases and the IV characteristics of both PN and LGAD devices become increasingly symmetric with respect to voltage polarity. This behavior is consistent with strong radiation-induced compensation of the originally N-type material, commonly attributed to the formation of deep acceptor-like defects.. The corresponding CV characteristics show that the bulk capacitance becomes nearly independent of bias voltage at high fluences, while remaining close to the geometrical value expected for the fully depleted epitaxial layer. This indicates that proton irradiation primarily affects the electrical properties of the epitaxial region without producing a pronounced expansion of the depletion depth into the substrate.

A qualitatively different behavior is observed after neutron irradiation. For neutron fluences up to \mbox{$2.3\times10^{17}\;\mathrm{n_{eq}/cm^2}$}, the CV characteristics remain broadly similar to those measured after proton irradiation, with the bulk capacitance after depletion staying close to the geometrical capacitance of the epitaxial layer. For the highest neutron fluences of \mbox{$5.2\times10^{17}$} and \mbox{$1.0\times10^{18}\;\mathrm{n_{eq}/cm^2}$}, however, the bulk capacitance decreases dramatically to approximately \mbox{2.2 pF}, corresponding to an effective depletion depth of about \mbox{350 $\mu$m}. This value significantly exceeds the nominal thickness of the epitaxial layer (\mbox{$50\;\mu$m}), indicating that the depletion region extends far beyond the epitaxial layer and penetrates deeply into the originally highly doped substrate. The extracted depletion depth corresponds to an effective space charge concentration on the order of \mbox{$10^{12}$--$10^{13}\;\mathrm{cm^{-3}}$}. Such a low effective doping indicates strong radiation-induced modification of the space charge in the irradiated material, leading to a substantial expansion of the depletion region even into regions that were originally highly doped.

The leakage current after neutron irradiation increases systematically with neutron fluence, reaching almost \mbox{10 nA} at \mbox{$1\times10^{18}\;\mathrm{n_{eq}/cm^2}$}. The large depletion depth inferred from the capacitance data indicates a substantial increase of the active volume contributing to the leakage current. However, the observed current level cannot be explained by simple bulk thermal SRH generation alone, given the extremely low intrinsic carrier concentration of 4H-SiC. This suggests that, in addition to the enlarged depletion volume, field-enhanced carrier generation and/or surface and edge-related leakage mechanisms also contribute to the measured current.

Irradiation by $^{60}$Co gamma rays up to a total ionizing dose of \mbox{300 kGy} produces only minor changes in the IV characteristics, with the reverse leakage current remaining statistically comparable to that of unirradiated devices. In contrast, the CV characteristics reveal a visible modification of the depletion behavior for the highest investigated dose. For PN diodes irradiated to \mbox{300 kGy}, the decrease of the bulk capacitance with reverse bias becomes significantly steeper, while the capacitance after full depletion remains close to the geometrical value of the epitaxial layer, reaching approximately \mbox{16.8 pF}. For LGAD devices irradiated to the same dose, the capacitance becomes nearly constant over the full range of reverse bias voltages, with values around \mbox{17 pF}. These observations indicate a reduction of the effective doping concentration in the epitaxial region and in the multiplication layer, while the effective thickness of the depleted region remains close to the original epitaxial thickness.

Overall, the presented results demonstrate the exceptional radiation hardness of 4H-SiC detectors and highlight the markedly different effects of proton, neutron, and gamma irradiation. While proton irradiation primarily modifies the electrical properties of the epitaxial layer, neutron irradiation can strongly modify the effective space charge and dramatically enlarge the depletion region, extending far beyond the nominal epitaxial layer thickness and deep into the substrate. In contrast, gamma irradiation predominantly affects the depletion behavior through ionization-related mechanisms without significantly changing the leakage current. These findings confirm the strong potential of 4H-SiC PN and LGAD sensors for operation in extreme radiation environments.

\section*{Acknowledgements}
\noindent This work was supported by the European Structural and Investment Funds and the Ministry of Education, Youth and Sports of the Czech Republic via the projects LM2023040 CERN-CZ and FORTE – CZ.02.01.01/00/22\_008/0004632. Financial resources were also provided by the Technological Agency of the Czech Republic through the project TK05020011.

Researcher Peter Švihra conducts his research under the Marie Skłodowska-Curie Actions – COFUND project Physics for Future (Grant Agreement No. 101081515), co-funded by the European Union. 

This project has received funding from the European Union’s Horizon 2020 research and innovation programme under Grant Agreement No. 654168.

The authors would like to thank the staff of the TRIGA Mark II reactor at the Jožef Stefan Institute in Ljubljana, the IRRAD facility at CERN, and the $^{60}$Co irradiation facility at UJP PRAHA, a.s., for their assistance with the irradiation of the samples used in this study.

\section*{Declaration of Generative AI and AI-assisted technologies in the writing process}
\noindent During the preparation of this work the authors used ChatGPT to improve the readability of the manuscript. After using this tool, the authors reviewed and edited the content as needed and take full responsibility for the content of the published article.

\section*{Declaration of competing interest}
\noindent The authors declare that they have no known competing financial interests or personal relationships that could have appeared to influence the work reported in this paper.




\bibliographystyle{elsarticle-num} 
\bibliography{kroll_bibliography.bib}



\end{document}